\documentclass[a4paper,preprintnumbers,amsmath,amssymb,twocolumn,10pt]{revtex4-2}

\usepackage{graphicx}
\usepackage{dcolumn}
\usepackage{bm}
\usepackage{epstopdf}
\usepackage{dsfont}
\usepackage{color}
\usepackage{amsthm}

\newcount\driver
\newcount\bozza

scaled\magstep1                  
scaled\magstep1 
 
scaled\magstep1                 \font\indbf=cmbx10 scaled\magstep2

scaled \magstep2               

{\count255=\time\divide\count255 by 60
\xdef\hourmin{\number\count255}
        \multiply\count255 by-60\advance\count255 by\time
   \xdef\hourmin{\hourmin:\ifnum\count255<10 0\fi\the\count255}}

\let\a=\alpha \let\b=\beta    \let\g=\gamma     \let\d=\delta     \let\e=\varepsilon
       \let\th=\vartheta \let\k=\kappa     \let\l=\lambda
    \let\n=\nu      \let\x=\xi                
\let\s=\sigma \let\t=\tau            
\let\ps=\psi   \let\o=\omega     
\let\G=\Gamma \let\D=\Delta       \let\L=\Lambda    
             
\let\O=\Omega

\def\RR{{\cal R}}\def\LL{{\cal L}}

\def\KK{{\cal K}}

\def\xx{{\bf x}}
  
\def\yy{{\bf y}}\def\kk{{\bf k}}\def\nn{{\bf n}}

       \def\oo{{\underline \omega}}
\def\ee{{\underline \varepsilon}}

\let\==\equiv

\let\io=\infty
\let\0=\noindent

\def\*{{\hfill\break\null\hfill\break}}

\def\tilde#1{{\widetilde #1}}

\def\tende#1{\,\vtop{\ialign{##\crcr\rightarrowfill\crcr
             \noalign{\kern-1pt\nointerlineskip}
             \hskip3.pt${\scriptstyle #1}$\hskip3.pt\crcr}}\,}
\def\otto{\,{\kern-1.truept\leftarrow\kern-5.truept\to\kern-1.truept}\,}

\def\wh#1{\widehat{#1}}
\def\hat#1{\wh{#1}}
\def\sqt[#1]#2{\root #1\of {#2}}

\def\bp{{\bar \ps}}

\def\RR{{\cal R}}\def\LL{{\cal L}}

\def\T#1{{#1_{\kern-3pt\lower7pt\hbox{$\widetilde{}$}}\kern3pt}}
\def\VVV#1{{\underline #1}_{\kern-3pt
\lower7pt\hbox{$\widetilde{}$}}\kern3pt\,}
\def\W#1{#1_{\kern-3pt\lower7.5pt\hbox{$\widetilde{}$}}\kern2pt\,}

\def\indica{\leaders \hbox to 0.5cm{\hss.\hss}\hfill}
\def\guida{\leaders\hbox to 1em{\hss.\hss}\hfill}
\mathchardef\oo= "0521

\def\xx{{\bf x}}
\def\yy{{\bf y}}\def\kk{{\bf k}}\def\nn{{\bf n}}

\def\oo{{\underline \omega}}

\def\qed{\raise1pt\hbox{\vrule height5pt width5pt depth0pt}}
 
  \def\bp{{\bar p}} 

\def\indic{\hbox{\raise-2pt \hbox{\indbf 1}}}

\def\ins#1#2#3{\vbox to0pt{\kern-#2 \hbox{\kern#1 #3}\vss}\nointerlineskip}

\newdimen\xshift \newdimen\xwidth \newdimen\yshift

\def\insertplot#1#2#3#4#5#6{\xwidth=#1pt \xshift=\hsize \advance\xshift by-\xwidth \divide\xshift by 2\begin{figure}[ht]
\vspace{#2pt} \hspace{\xshift}
\begin{minipage}{#1pt}
#3 \ifnum\driver=1 \griglia=#6
\ifnum\griglia=1 \openout13=griglia.ps \write13{gsave .2
setlinewidth} \write13{0 10 #1 {dup 0 moveto #2 lineto } for}
\write13{0 10 #2 {dup 0 exch moveto #1 exch lineto } for}
\write13{stroke} \write13{.5 setlinewidth} \write13{0 50 #1 {dup 0
moveto #2 lineto } for} \write13{0 50 #2 {dup 0 exch moveto #1
exch lineto } for} \write13{stroke grestore} \closeout13
\includegraphics{griglia.ps} \fi
\includegraphics{#4.ps}\fi\ifnum\driver=2 \fi
\end{minipage}
\caption{#5}
\end{figure}
}

\newdimen\shift \shift=-1.5truecm
\def\lb#1{\ifnum\bozza=1
\label{#1}\rlap{\hbox{\hskip\shift$\scriptstyle#1$}}
\else\label{#1} \fi}

\def\be{\begin{equation}}
\def\ee{\end{equation}}
\def\bea{\begin{eqnarray}}\def\eea{\end{eqnarray}}
\def\bean{\begin{eqnarray*}}\def\eean{\end{eqnarray*}}
\def\bfr{\begin{flushright}}\def\efr{\end{flushright}}
\def\bc{\begin{center}}\def\ec{\end{center}}
\def\bal{\begin{align}}\def\eal{\end{align}}
\def\ba#1{\begin{array}{#1}} \def\ea{\end{array}}
\def\bd{\begin{description}}\def\ed{\end{description}}
\def\bv{\begin{verbatim}}\def\ev{\end{verbatim}}
\def\nn{\nonumber}
\def\Halmos{\hfill\vrule height10pt width4pt depth2pt \par\hbox to \hsize{}}
\def\pref#1{(\ref{#1})}

\def\ins#1#2#3{\vbox to0pt{\kern-#2 \hbox{\kern#1 #3}\vss}\nointerlineskip}

\newdimen\xshift \newdimen\xwidth \newdimen\yshift
\newcount\griglia

\def\insertplot#1#2#3#4#5#6{\xwidth=#1pt \xshift=\hsize \advance\xshift by-\xwidth \divide\xshift by 2\begin{figure}[ht]
\vspace{#2pt} \hspace{\xshift}
\begin{minipage}{#1pt}
#3 \ifnum\driver=1 \griglia=#6
\ifnum\griglia=1 \openout13=griglia.ps \write13{gsave .2
setlinewidth} \write13{0 10 #1 {dup 0 moveto #2 lineto } for}
\write13{0 10 #2 {dup 0 exch moveto #1 exch lineto } for}
\write13{stroke} \write13{.5 setlinewidth} \write13{0 50 #1 {dup 0
moveto #2 lineto } for} \write13{0 50 #2 {dup 0 exch moveto #1
exch lineto } for} \write13{stroke grestore} \closeout13
\includegraphics{griglia.ps} \fi
\includegraphics{#4.ps}\fi\ifnum\driver=2 \fi
\end{minipage}
\caption{#5}
\end{figure}
}

\newdimen\shift \shift=-1.5truecm
\def\lb#1{\label{#1}\rlap{\hbox{\hskip\shift$\scriptstyle#1$}}
\else\label{#1} \fi}

\def\be{\begin{equation}}
\def\ee{\end{equation}}
\def\bea{\begin{eqnarray}}\def\eea{\end{eqnarray}}
\def\bean{\begin{eqnarray*}}\def\eean{\end{eqnarray*}}
\def\bfr{\begin{flushright}}\def\efr{\end{flushright}}
\def\bc{\begin{center}}\def\ec{\end{center}}
\def\bal{\begin{align}}\def\eal{\end{align}}
\def\ba#1{\begin{array}{#1}} \def\ea{\end{array}}
\def\bd{\begin{description}}\def\ed{\end{description}}
\def\bv{\begin{verbatim}}\def\ev{\end{verbatim}}
\def\nn{\nonumber}
\def\Halmos{\hfill\vrule height10pt width4pt depth2pt \par\hbox to \hsize{}}
\def\pref#1{(\ref{#1})}

\usepackage{amsmath}
\usepackage{amsfonts}
\usepackage{amssymb}
\usepackage{epstopdf}

scaled\magstep1

\let\a=\alpha \let\b=\beta  \let\g=\gamma  \let\d=\delta
\let\e=\varepsilon
     \let\th=\theta \let\k=\kappa \let\l=\lambda
    \let\n=\nu    \let\x=\xi         
\let\s=\sigma \let\t=\tau    
\let\ps=\Psi   \let\o=\omega
\let\G=\Gamma \let\D=\Delta  \let\L=\Lambda 
         
\let\O=\Omega

\def\RR{{\cal R}}\def\LL{{\cal L}}  
 
\def\KK{{\cal K}}

 \def\xx{{\bf x}} \def\yy{{\bf y}} 
\def\kk{{\bf k}}

\def\nn{\nonumber}

\def\\{\hfill\break}
\def\={:=}
\let\io=\infty
\let\0=\noindent

\def\tende#1{\,\vtop{\ialign{##\crcr\rightarrowfill\crcr\noalign{\kern-1pt
    \nointerlineskip} \hskip3.pt${\scriptstyle #1}$\hskip3.pt\crcr}}\,}
\def\otto{\,{\kern-1.truept\leftarrow\kern-5.truept\to\kern-1.truept}\,}

\def\wh{\widehat}
\def\to{\rightarrow}

\def\qed{\hfill\raise1pt\hbox{\vrule height5pt width5pt depth0pt}}

\def\be{\begin{equation}}
\def\ee{\end{equation}}
\def\bp{\begin{pmatrix}}
\def\ep{\end{pmatrix}}
\def\bea{\begin{eqnarray}}
\def\eea{\end{eqnarray}}
\def\nn{\nonumber}
\def\pref#1{(\ref{#1})}

\def\lb{\label}

\newtheorem{lemma}{Lemma}[section]
\newtheorem{remark}{Remark}[section]
\newtheorem{theorem}{Theorem}[section]

\begin{document}

\title{Incommensurate Twisted Bilayer Graphene: emerging quasi-periodicity and stability}
 
\author{Ian Jauslin}
\affiliation{Rutgers University, Department of Mathematics, New Brunswick, USA}
\email{ian.jauslin@rutgers.edu}

\author{Vieri Mastropietro}
\affiliation{Università di Roma ``La Sapienza'', Department of Physics, Rome, Italy }
\email{vieri.mastropietro@uniroma1.it}
 
\begin{abstract} 
We consider a lattice model of twisted bilayer graphene (TBG) for incommensurate twist angles, focusing on the role of large-momentum-transfer Umklapp terms. These terms, which nearly connect the Fermi points of different layers, are typically neglected in effective continuum descriptions but could, in principle, destroy the Dirac cones;
they are indeed closely analogous to those appearing in fermions within quasi-periodic potentials, where they play a crucial role.
We prove that, for small but finite interlayer coupling, the semimetallic phase is stable 
provided the angles belong to a fractal set of large measure (which decreases with the hopping strength)
characterized by a number-theoretic Diophantine condition. In particular, this set excludes the (zero measure) commensurate angles.
Our method combines a Renormalization Group (RG) analysis of the imaginary-time, zero-temperature Green's functions, with number theoretic
properties, and it is similar to the technique used in the Lindstedt series approach to Kolmogorov-Arnold-Moser (KAM) theory. The convergence of the resulting series allows us to rule out non-perturbative effects. The result provides a partial justification
of the effective continuum description of TBG in which such large-momentum interlayer hopping processes
are neglected.
\end{abstract} 
\maketitle

\section{Introduction}  

{\it Twisted Bilayer Graphene (TBG) } consists of two graphene layers, stacked and twisted by an angle $\th$.
The electronic properties of graphene are accurately captured
by a tight-binding model of electrons hopping from one site to a neighboring one
of a honeycomb lattice; the corresponding energy
bands can be explicitly computed and
intersect at two Dirac points, close to which the effective dispersion relation is approximately conical.
In the case of half-filling, low-energy excitations have an effective continuum description in terms of two-dimensional massless 
Dirac fermions and the system behaves as a Dirac semimetal.

Stacking and twisting two honeycomb lattices produces an intricate Moir\'e pattern
\cite{a1} and twisted bilayer graphene behaves very differently from the single layer, with a strong sensitivity to the twist angle,  
ranging from non-conductive to superconductive \cite{a}, see e.g. \cite{b},\cite{ba},\cite{bab}.
From the theoretical side, in contrast to the monolayer case, 
an exact computation of the dispersion relation is impossible and 
the validity of effective models is less clear.

The model for TBG we will consider consists of two layers of graphene with a quadratic interlayer coupling $\l$ smaller than the intralayer hopping (chosen equal to $1$).
We denote the twist angle by
$\th$. One has to distinguish between {\it commensurate} angles, such that the TBG
is periodic and which are
a discrete zero measure set indexed by two integers 
\cite{1bb},\cite{1aaa},\cite{H5}; and  {\it incommensurate} angles
where periodicity is lost and Bloch theory cannot be applied \cite{P0}.

The Dirac points of layer $1$ are $K_\pm={2\pi\over 3}(1,\pm  {\textstyle{1\over \sqrt{3}}})$
and the reciprocal lattice vectors are $b_1, b_2$; correspondingly the Dirac points of layer 2 are $K^{'}_\pm=R(\theta) K_\pm$ in which $R(\th)$ is the rotation matrix of angle $\theta$, and the reciprocal lattice vectors are $b^{'}_i=R(\theta) b_i$, $i=1,2$. The interlayer hopping with coupling $\l$ produces a sum of terms with 
an 
exchange of momenta given by multiples of the reciprocal lattice vectors; if $k_a,k_b$ are the momenta of two fermions, the conservation rule is given by
\begin{equation}\label{sss}
 k_a-k_b+G+G'=0
\end{equation}
 with $G=l_1b_1+l_2b_2= lb$, $G'=m_1b'_1+m_2b_2'= mb'$,
$l=(l_1,l_2)$, $m\equiv(m_1,m_2)$ with $l_i,m_i$ integers. 
The processes involving non-vanishing $G+G'$ are called 
{\it Umklapp terms}. 
The interlayer hopping decreases with large exchange of momentum. 

\begin{figure}
\includegraphics[width=3cm]{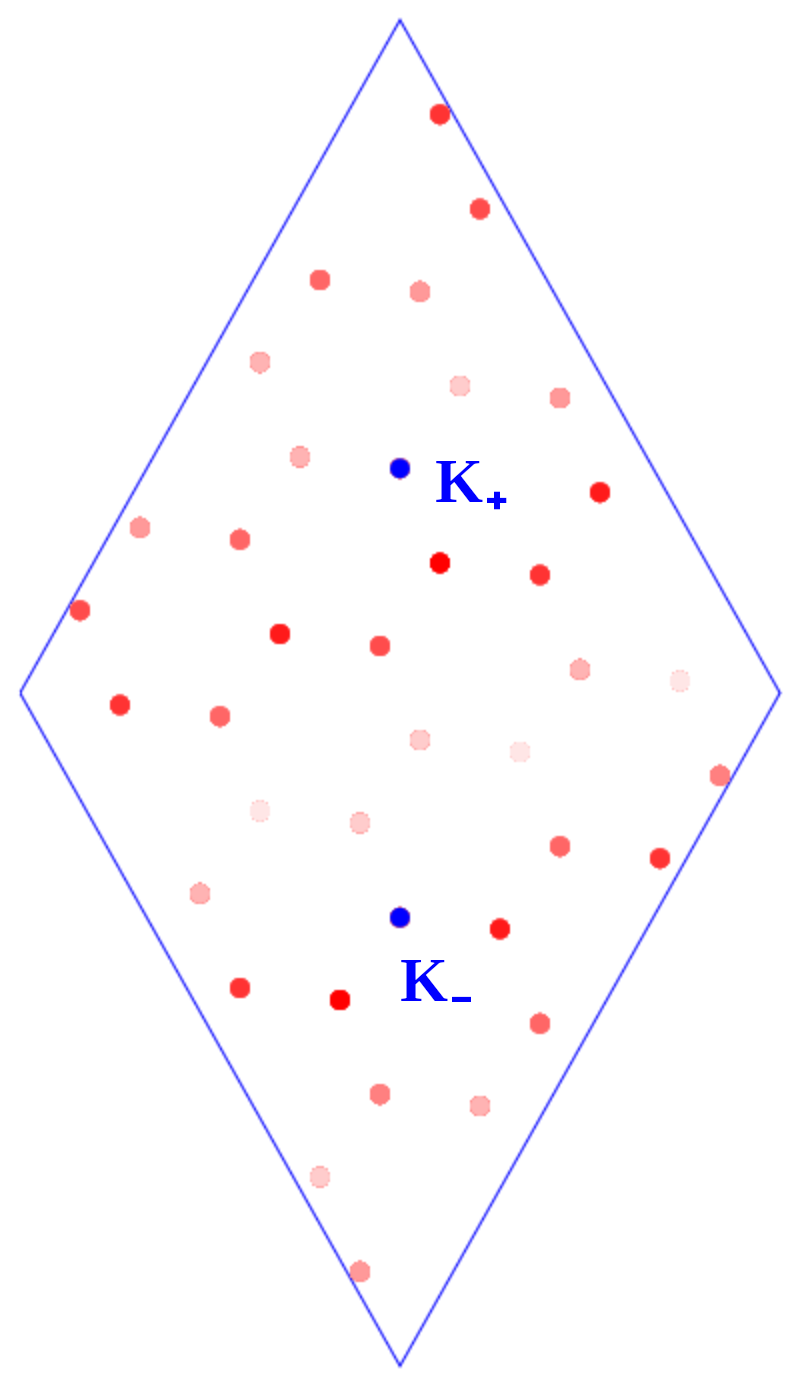}
\hskip1cm
\includegraphics[width=3cm]{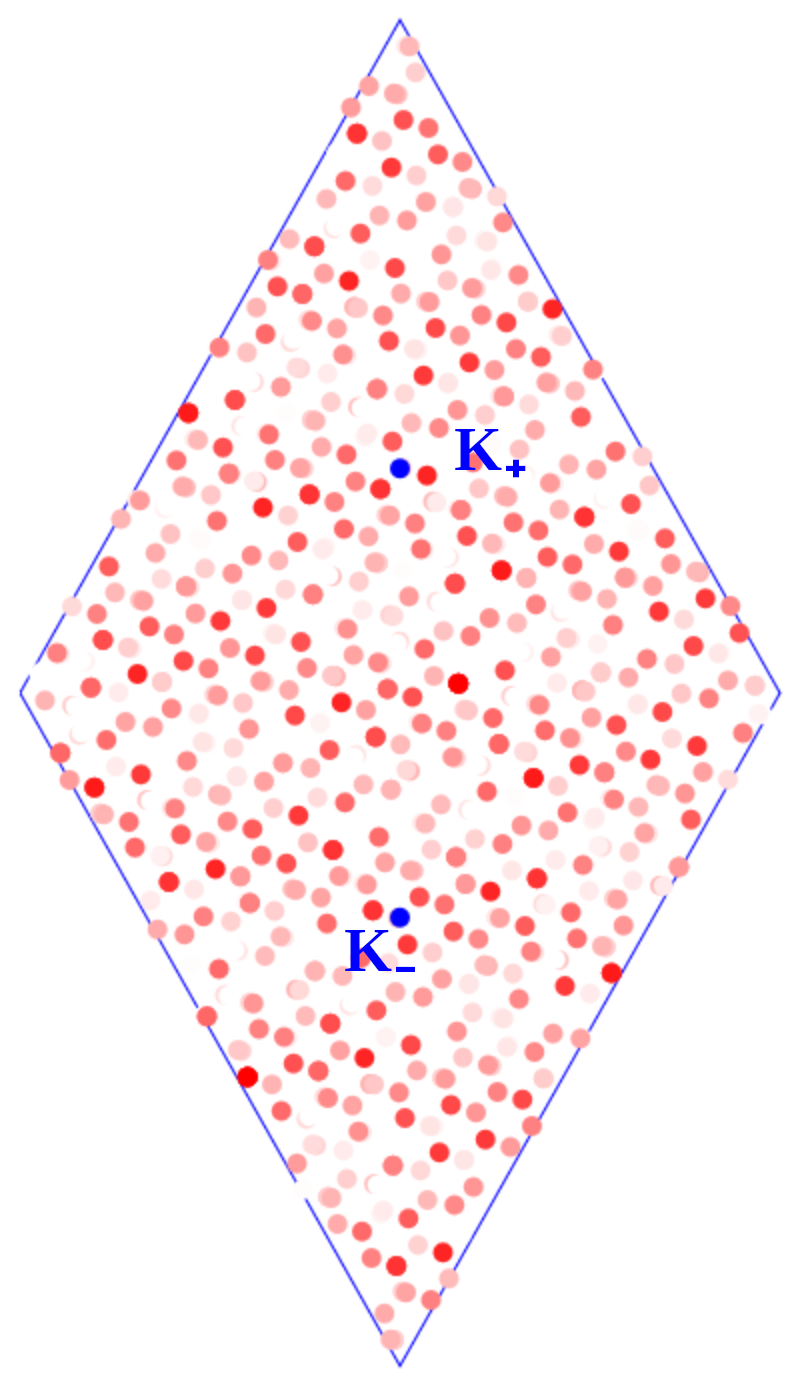}
\caption{\label{fig:feynman1} 
Left: The Dirac points in the commensurate case $\cos \th=37/38$.
Right: The Dirac points in the approximately incommensurate case $\theta=0.16\pi$.
The Dirac points $K_\pm$ of one layer are in blue (color online), and those
of the other layer $K_\pm'+G+G'$ are in red.
In the incommensurate case, $K'_\pm+G+G'$ get arbitrarily close to $K_\pm$.
}
\end{figure}

The processes connecting the Dirac points are of particular importance, that is the processes such that
$\Delta=0$ where
\begin{equation}\label{fgf}
\Delta=|G+G'+ K_\s^i-K_{\s'}^j|\equiv|M+ K_\s^i-K_{\s'}^j |
\end{equation}
with $K_\pm^1=K^\pm$, $K_{\pm}^2=K^{'}_\pm$ and $M=G+G'$. 
Among such terms, one can distinguish between the ones
connecting the same Dirac point (that is $i=j,\s=\s'$) from those connecting different Dirac points.
The form of the former is determined by the lattice symmetries 
\cite{1aa1,1aa2} and their effect is to renormalize the parameters of the single layer Hamiltonian, possibly enlarged to include new terms allowed by the breaking of symmetries produced by the interlayer hopping.
The latter, that is the ones involving different Dirac points,
are present only for {\it commensurate} angles and are expected to be relevant and able to destroy the Dirac cones \cite{112}.
One can distinguish between the intralayer terms ($i=j,\s\not =\s'$) and interlayer terms ($i\not =j,\s\not =\s'$);
the first case corresponds to the opening of a gap and the other to quadratic bands.
The gap is expected to decrease with the momentum transferred due to the decay in momentum of the hopping.
Due to periodicity there is only a finite set of exchanged Umklapp momenta, see Figure \ref{fig:feynman1}.
Therefore, in general, in the commensurate case the terms that do not connect the Dirac points exactly have a large mismatch and are expected to be irrelevant.

Also the terms {\it almost} connecting the Dirac points can produce an important modification
of the physical properties with respect to graphene. Terms almost connecting  the Dirac points can be of two kinds.
A first possibility is for angles $\th$ close to zero, where one can see that $\D$ can be $O(\theta)$; they are present both in the commensurate or incommensurate case.
An effective model was proposed in 
\cite{1},\cite{2}, \cite{3} keeping only a few terms 
($G=G'=0$, or  $G=b_1, G'=-b'_1$ or $G=b_2, G'=-b'_2$) corresponding to $\D=O(\theta)$
and passing to the continuum. It was found that the effective hopping is greatly enlarged: it becomes of order
$O(\l/\th)$; therefore the conical dispersion relation can persist 
but with a strongly decreased velocity; for certain "magic" angles the velocity even vanishes and,
remarkably, superconductivity has been found \cite{a}.
Effective models of TBG in which not only the velocity vanishes but the bands are completely flat have been also proposed
\cite{3a} and mathematical derivations of such effective models in suitable limits have been provided \cite{3b}.

The other possibility is present only in the incommensurate case and involves large momentum transfer.
Indeed in this case there are infinitely many Umklapp terms
{\it almost} connecting the Dirac points, that is such that
the distance between $K_\pm$ and 
$K_\pm'+G'+G$ 
can be arbitrarily small, that is $\Delta$ can be arbitrarily small, see figure \ref{fig:feynman1}.
These terms, which nearly connect the Fermi points of different layers, are typically neglected in effective continuum descriptions but could, in principle, destroy the Dirac cones;
they are indeed closely analogous to those appearing in the case of fermions in quasi-periodic potentials, where they play a crucial role.
This can be easily realized 
comparing TBG for incommensurate angles with fermions on a chain with step $1$ and with 
an Aubry quasi-periodic potential $\cos 2\pi \o x$ \cite{Au}, whose single partcle properties are described by the Harper equation. In that case the conservation of momentum reads
$k_a-k_b+2l\pi+2\pi \o m=0$ and the mismatch between Fermi points is  
\be 
\D=|2 \e K+2l\pi+2\pi \o m| \quad \e=0,\pm 1
\ee
if $K$ is the Fermi momentum. If $\o/2\pi$ is rational either the above quantity is zero or is bounded while if 
it is irrational there are almost relevant processes that almost connect the Fermi points. The quasi-periodicity acts as a disorder and in the strong coupling regime there
is an insulating localized behaviour while in the weak coupling a metallic behaviour is found, as proved in 
\cite{DS}, \cite{FS}, \cite{A}. The spectrum has 
remarkable fractal properties, which were observed in experiments with magnetic fields \cite{K}. 

The prototypical problem involving quasi-periodicity
is the Kolmogorov-Arnold-Moser (KAM) theorem for the persistence of Hamiltonian invariant tori under perturbation, and, in fact, results on
the Aubry potential use KAM methods. 
The Renormalization Group can be used to prove the convergence of the KAM Lindstedt series
\cite{G} or to compute the Green's function of 
fermions in a quasi periodic potential \cite{M10}. It is also known that the interplay of many body interaction and quasi-periodic potential
leads to interesting phenomena, studied in $1d$ fermions in the delocalized regime in  
\cite{M11}, \cite{Q0} and in the localized one in \cite{Q2}, \cite{M12}. Quasi-periodic potentials have also been investigated in interacting Weyl semimetals
\cite{P4}, \cite{W1}. Note also that
the large-momentum Umklapp terms in incommensurate TBG are also present when considering a hopping with 
a finite number of harmonics; they would be generated at higher order in perturbation theory.
This fact is not captured by the effective models, as the lattice is replaced by the continuum and Umklapp terms are therefore absent.
These large-momentum Umklapp terms are depressed by the decay of the hopping in momentum space, but there is no guarantee that this makes them irrelevant; in the case of the Aubry potential they indeed produce an Anderson localized phase.

In order to evaluate the effect of the Umklapp
large-momentum terms on the physical properties of the system we need, as is the case in KAM theory or for quasi-periodic systems, to use
number theoretic properties of irrationals. Such properties are typically established in the context of measure theory.
The typical number theoretic condition which is assumed in the case of problems involving quasi-periodicity is 
a {\it Diophantine condition}
which consists in restricting ourselves to values of $\theta$ for which, for any $\o, \o', i, j$,
\be
|K^i_\o - K^j_{\o'} + l b + m b'| \ge \frac{C_0}{|y|^\t}
\label{cond}
,
\ee
where $y$ is either $y=l$ or $y=m$ and $y\neq0$; this relates the 
size of the momentum $G,G'$ with the precision with which the Dirac points are connected. 
A similar Diophantine condition is assumed in the case of Aubry-André systems, where the condition is on the frequency $\o$, and in KAM theory where the condition is on the frequency of the tori. $C_0$ is called the Diophantine constant and it plays a crucial role.

We consider a lattice model of TBG and we perform an exact Renormalization Group analysis
of the 2-point zero temperature imaginary time Green function in order to find conditions for the 
persistence of the Dirac cones and the stability of the semimetallic phase. Such zero temperature correlations provide information on the ground state and in particular on the persistence or not of the semimetallic behaviour. Note that, as we are not considering a many-body interaction, information
on the cones could also be obtained, in principle, by studying the corresponding one-particle Schrödinger operator. We find however
more suitable the functional integral formalism, considering the hopping as a perturbation and expanding
the correlation as a convergent series. 

The stability is a non-perturbative phenomenon which cannot be understood on the basis 
of lowest order computations but it depends on the convergence or divergence of the whole series. This is a common feature in problems involving 
incommensurability; the KAM series are convergent and correspond to stable motions while the Birkhoff series are diverging and correspond to chaos but both have a
small divisor problem.

We choose an interlayer hopping with coupling $\lambda$ that decays exponentially for large momenta and satisfies suitable lattice symmetries so that the terms connecting the same Dirac points just renormalize the single layer parameters
except for a modification of the location of Dirac points. Such a choice of symmetries is done for definiteness and it could be avoided 
by including other terms (counterterms) in the single layer Hamiltonian that are compatible with the symmetries, while the fast decay of the hopping in momentum space seems essential.

Our main result is the following:
\vskip.2cm
{\it In a lattice model of twisted bilayer graphene with an angle $\theta$ verifying
the Diophantine condition \pref{cond} for 
$|\l|\le \e_0(C_0)$ with $\e_0(C_0)$ proportional to $C_0^\beta$ for some positive power $\beta$, the Dirac cones persist and 
the Dirac semimetallic phase is stable.} 
\vskip.2cm
This result ensures that the large momentum Umklapp terms, even if they almost connect the Dirac points, are not able
to destroy the semimetallic phase at weak coupling.
The proof is based on  
a Renormalization Group analysis for the zero temperature imaginary time Green functions and
the convergence of the series allows to exclude non-perturbative effects.

The stability of the cones is established with a smallness condition which depends on a power of the Diophantine constant $C_0$;
given an angle one can in principle compute it but it is expected to depend wildly on the exact value of $\theta$. 
It is therefore convenient to give a generic statement for angles in a certain set, as it is done for disordered systems where the properties
hold for generic (not all) configurations of disorder.
Given an angle in a certain interval
of size $2\,\delta\theta$ around $\theta$, we can prove that the angles verifying \pref{cond}
have almost a full measure: the complement of the set has a measure that is bounded by $C_{\theta,\delta\theta} C_0^\alpha(2\,\delta \theta)$, with some positive power $\alpha$ and 
a constant $C_{\theta,\delta\theta}$ (that depends on
derivatives of $K^\pm_i, b_j$); 
the explicit value of the power $\alpha$ and the form of $C_{\theta,\delta\theta}$
are written explicitly below. One can check that $C_{\theta,\delta\theta}$ is finite apart from a finite set of special, pathological angles (there are 20 such pathological angles, which arise from technical issues). Therefore choosing a hopping $\lambda$ fixes an admissible value for $C_0$, hence choosing, say, $\lambda=\e_0(C_0)/2$
the measure of the angles for which the Dirac cones persist has a complement that is bounded by
$C_{\theta,\delta\theta} C_0^\alpha (2\, \delta \theta)$. In addition $\d \th$ also has to be chosen small with respect to $C_0$ but this is probably just a technical assumption which could be removed by a lengthier analysis.
The set for which stability is ensured is of course rather complex and has a fractal nature and its relative measure decreases with the hopping.
The situation is therefore entirely analogous to the breaking of invariant tori in KAM theory,
see e.g.  \cite{K0}, \cite{K1}.
Note that the excluded set contains the commensurate angles, but 
experimentally to see an instability can be difficult as the possible
gap could be very small if the corresponding Umklapp momentum is very large.
Another equivalent way to state the result is that for any angle (except the pathological ones) one can find an interval of angles near it such that,
with probability $1$, there is a non-zero value of the hopping for which the
semimetallic behaviour persists.

Our result therefore ensures that for weak hopping for most (in the sense specified above) angles the semimetallic behaviour persists
even in presence of the large momentum Umklapp terms which almost connect the Dirac points.
The result provides a partial justification
of the effective continuum description of TBG in which terms are neglected. It would be of course a very interesting problem
to consider hopping outside the regime we are considering to see if there can be some transition to a different phase due to such terms,
as it happens for fermions in a 1D quasi-periodic Aubry potential; or to detect in the weak coupling regime some sign of fractality 
following from the above number theoretical considerations.

The paper is organized in the following way. In Section \ref{sec:model} the lattice model of TBG
is presented. In Section \ref{sec:feynman} a perturbative expansion for the correlations is derived.
The main result is stated in Section \ref{sec:result}.
In Section \ref{sec:feynman1} the emerging quasi-periodicity and the small divisor problem is described, together with the required (number theoretic) Diophantine conditions. In Section \ref{sec:renormalized} the 
Renormalization Group derivation is presented. The Appendices detail the more technical aspects of the analysis.

\section{Incommensurate TBG}\label{sec:model}

We consider the lattice TBG model introduced in \cite{1,2}.
We focus on this model for the sake of definiteness but our methods could be applied more generally.
We consider two graphene layers rotated with respect to one another by an angle $\theta$ around a point $\xi=(0,1/2)$ (that is, the point between an a and b atom, chosen so that the twisted model preserves the $C_2T$ symmetry in Appendix \ref{sec:symmetry}).
The Hamiltonian of the system will be written as
\begin{equation}
  H=H_1+H_2+V
\end{equation}
where $H_1$ and $H_2$ are hopping Hamiltonians within the layers 1 and 2 respectively and $V$ is an interlayer hopping term.
The first graphene layer is defined on the
lattice $\mathcal L_1:=\{n_1 A_1+n_2 A_2,\ n_1,n_2\in\mathbb Z\}$
with $A_1={1\over 2}(3,\sqrt{3})
  ,\quad
  A_2={1\over 2}(3,-\sqrt{3})$.
We introduce the nearest-neighbor vectors: $\d_1=(1,0)$, $\d_2={1\over 2}(-1, \sqrt{3})$, $\d_2={1\over 2}(-1, -\sqrt{3})$.
We will write the Hamiltonian in second quantized form: for $x\in \mathcal L_1$, we introduce {\it annihilation operators} $c_{1,x,a}$ and $c_{1,x,b}$ corresponding respectively to annihilating a fermion located at $x$ and $x+\d_1$.
The nearest neighbor hopping Hamiltonian is ($t=1$ from now on)
\be H_1= -t\sum_{x\in \mathcal L_1}\sum_{i=0}^2 (c_{1,x,a}^\dagger c_{1,x+A_i,b}+ c_{1,x+A_i,b}^\dagger c_{1,x,a})\ee
where $A_0:=0$ (note that $\d_1-\d_2=A_2, \d_2-\d_3=A_3$).
We will do much of the computation in Fourier space, and we here introduce the Fourier transform $\hat c_{1,k,\alpha}$ of $c_{1,x,\alpha}^\pm$ in such a way that, for $\alpha\in\{a,b\}$,
\begin{equation}
  c_{1,x,\alpha}
  =\frac1{|\hat{\mathcal L}_1|}\int_{\hat{\mathcal L}_1} dk\ e^{-ik(x-\xi)}\hat c_{1,k,\alpha}
\end{equation}
with  $|\hat{\mathcal L}_1|=8 \pi^2/3 \sqrt{3}$, and
$
  \hat{\mathcal L}_1:=\mathbb R^2/(b_1\mathbb Z+b_2\mathbb Z)$ in which
$b_1= {\textstyle{2\pi\over 3}}(1,\sqrt{3})
 ,\quad
 b_2= {\textstyle{2\pi\over 3}}(1,-\sqrt{3})$
In Fourier space,
\begin{equation}
  H_1= 
\frac t{|\hat{\mathcal L}_1|}\int_{\hat{\mathcal L}_1}dk\ \left(\Omega(k)\hat c_{1,k,a}^\dagger\hat c_{1,k,b}+\Omega^*(k)\hat c_{1,k,b}^\dagger\hat c_{1,k,a}\right)
\label{H1k}
\end{equation}
with $\Omega(k_x,k_y):=1+2e^{-i\frac32k_x}\cos({\textstyle\frac{\sqrt 3}2k_y})$.
Note that, for small $\th$, $|K_\pm-K^{'}_\pm|=O(\theta)$.
.

The second graphene layer is rotated by an angle $\theta$ around the point $\xi=(0,1/2)$, that is, it is defined on the lattice
\be \mathcal L_2=\xi+R(\theta)(\mathcal L_1-\xi),\quad R(\th)=\begin{pmatrix} c_\th &-s_\th \\s_\th & c_\th \end {pmatrix}\ee
(we use the shorthand throughout this paper that $c_\th\equiv\cos\th, s_\th\equiv\sin \th$).
The annihilation operators in the second layer are denoted by $c_{2,x,a}$ and $c_{2,x,b}$.
The hopping Hamiltonian of this second layer is
\begin{equation}
  H_2= -t\sum_{x\in \mathcal L_2}\sum_{i=0}^2 (c_{2,x,a}^\dagger c_{2,x+RA_i,b}+ c_{2,x+R A_i,b}^\dagger c_{2,x,a})
\end{equation}
where $R\equiv R(\theta)$.
We define the Fourier transform in the second layer: if
$b'_1:=R b_1$, $b'_2:=R b_2$
and
\begin{equation}
  c_{2,x,\alpha}
  =\frac1{|\hat{\mathcal L}_1|}\int_{\hat{\mathcal L}_2} dk\ e^{-ik(x-\xi)}\hat c_{2,k,\alpha}
\end{equation}
we find
\begin{equation}
\begin{array}{r@{\ }>\displaystyle l}
H_2=&
\frac t{|\hat{\mathcal L}_1|}\int_{\hat{\mathcal L}_2}d k\ 
\cdot\\&\cdot\left(\Omega(R^T k)\hat c_{2,k,a}^\dagger\hat c_{2,k,b}+\Omega^*(R^Tk)\hat c_{2,k,b}^\dagger\hat c_{2,k,a}\right)
.
\label{H2k}
\end{array}
\end{equation}

In the absence of interlayer coupling the two graphene layers are decoupled;
the single particle spectrum for layer 1 is $\pm |\Omega(k)|$
and the Fermi points
are given by the relation
$\O(K^1_\pm)=0$ with
\begin{equation}
  K^1_\pm={2\pi\over 3}(1,\pm  {\textstyle{1\over \sqrt{3}}})
  \label{pF}
\end{equation}
for momenta close to such points one has 
$|\Omega(k)| \sim {3\over 2} t  |k-K^1_\pm|$, that is the dispersion relation is 
almost linear (relativistic) up to quadratic corrections, forming approximate {\it Dirac cones}. In the same way the 
dispersion relation for layer 2 is $\pm |\Omega(R^T k)|$; 
the Fermi points are $\O(R^T K^2_\pm)=0$ with
$K^2_\pm=R(K^1_\pm)$
and $|\Omega(R^T k)| \sim {3\over 2} t  |k-K^2_\pm|$. We are interested in understanding how these four Dirac cones are modified in the presence of the interlayer hopping. 

We couple the 2 layers by an interlayer hopping Hamiltonian, which couples atoms of type a to atoms of type b:
\begin{equation}
\begin{array}{>\displaystyle l}
V=
\l \sum_{x_1\in \mathcal L_1} \sum_{x'_2\in\mathcal L_2}\sum_{\alpha\in\{a,b\}} \varsigma(x_1+d_\alpha-x'_2-Rd_{\alpha})
\cdot\\
\hfill\cdot(c^\dagger_{1,x_1,\alpha}c_{2,x'_2,\alpha}+c_{2,x_2',\alpha}^\dagger c_{1,x_1,\alpha})
\end{array}\end{equation}
$d_a=(0,0), d_b=\d_1$, and $\varsigma(x)=\varsigma(-x)$,
\be
\varsigma(x_1-x_2)=\int_{\mathbb R^2}\frac{dq}{4\pi^2}\  e^{i q(x_1-x_2)} \hat \varsigma(q)
,\quad
|\hat\varsigma(q)|\le e^{-\k |q|}
.
\label{interlayer}
\ee
We restrict the interlayer term to hoppings between atoms of type a to atoms of type a and type b to b so that
the ``Inversion'' symmetry in Appendix \ref{sec:symmetry} is satisfied.
Note that, whereas the Fourier transform for $c$ is defined on $\hat{\mathcal L}_i$, the Fourier transform of $\varsigma$ is defined on all $\mathbb R^2$.
We write $V$ in Fourier space: we get, see App. \ref{app:fourierV}
\bea
&&V=
\frac \l{4\pi^2|\hat{\mathcal L}_1|}\sum_{\alpha}\left(\sum_{l\in\mathbb Z^2}\int_{\hat{\mathcal L}_1}d k \ 
\tau^{(1)}_{l,\alpha}(k+l b)
\hat c^\dagger_{1,k,\alpha}\hat c_{2,k+l b,\alpha}
\right.\nn\\
&&\left.+\sum_{m\in\mathbb Z^2}\int_{\hat{\mathcal L}_2}d k\
\tau_{m,\alpha}^{(2)}(k+m b') 
\hat c^\dagger_{2,k,\alpha}\hat c_{1,k+m b',\alpha}\right)
\label{11} 
\eea
where we use the notation $lb\equiv l_1b_1+l_2b_2$, $mb'\equiv m_1b_1'+m_2b_2'$, and
\begin{equation}
\tau^{(1)}_{l,\alpha}(k)
  :=e^{i \xi lb}e^{-ik(d_\alpha-Rd_{\alpha})}
  e^{-i\xi \sigma_{k,1}b'}\hat\varsigma^*(k)
  \label{tau1}
  \end{equation}
\begin{equation}
\tau^{(2)}_{m,\alpha}(k):=e^{i\xi mb'}e^{-ik(d_\alpha-Rd_{\alpha})}e^{-i\xi \sigma_{k,2}b}\hat\varsigma(k)
  \label{tau2}
\end{equation}
in which $\sigma_{k,i}\in \mathbb Z^2$ is the unique integer vector such that
$
  k-\sigma_{k,1}b'\in\hat{\mathcal L}_2
  ,\ 
  k-\sigma_{k,2}b\in\hat{\mathcal L}_1
  $. Note that the difference of the momenta of the two fermions is given by $l b+m b'$.

The position of the Dirac points are in general modified (renormalized)
by the interlayer hopping.
It is convenient to fix the values of the renormalized
Dirac points by properly choosing the bare ones. This can be achieved by replacing
$\O(k)$
with $\O(k)+\nu_{i,\o}$ close to each Dirac points, that is adding
a counterterm has the form
\begin{eqnarray}
&&M=
  \sum_{\omega\in\{+,-\}}\sum_{i=1,2}\int_{\hat{\mathcal L}_i} dk\ \chi_{\omega,i}(k)(  \nu_{i,\omega} \hat c_{i,k,a}^\dagger\hat c_{i,k,b}\nn\\
&&+ \nu^*_{i,\omega} \hat c_{i,k,b}^\dagger\hat c_{i,k,a})
\label{counterterms}
\end{eqnarray}
where $\chi_{\o,i}(k)$ is a smooth compactly supported function  that is
non vanishing for  $||k-K^i_\o||_i\le 1/\gamma$, for some $\gamma>1$, in which $||.||_i$ is the norm on the torus $\hat{\mathcal L}_i$.

Our main result concerns the Matsubara imaginary time Green functions, which we define as follows.
We first introduce a Euclidean time component: given an inverse temperature $\beta>0$, we define for $x_0\in[0,\beta)$,
\begin{equation}
  c_{j,x,\alpha}(x_0):=e^{-x_0\bar H}c_{j,x,\alpha}e^{x_0\bar H}
  .
\end{equation}
and $\bar H=H+M$. Combining the Euclidean time component with the spatial one, we define $\Lambda_i:=[0,\infty)\times \mathcal L_i$.
The corresponding Fourier-space operators are
\begin{equation}
  \hat
  c_{j,k,\alpha}(k_0)=
  \int_0^\beta dx_0 e^{-ix_0k_0}\sum_{x\in \mathcal L_j}e^{-i(x-\xi)k}c_{j,x,\alpha}(x_0)
\end{equation}
which is defined for $(k_0,k)\in\hat \Lambda_j:=\frac{2\pi}\beta (\mathbb Z+1/2)\times \hat{\mathcal L}_j$.

Now, given $j,j'\in\{1,2\}$, $\mathbf k=(k_0,k)\in \Lambda_j$, the 2-point function is defined as the $2\times2$ matrix $\hat S_{j,j'}(\mathbf k)$ whose components are indexed by $\alpha,\alpha'\in\{a,b\}$:
\begin{equation}\label{xx}
  (\hat S_{j,j'}(\mathbf k))_{\alpha,\alpha'}:=
  \frac{\mathrm{Tr}(e^{-\beta \bar H}T(\hat c_{j,k,\alpha}(k_0),\hat c_{j',k,\alpha'}^\dagger(k_0)))}{\mathrm{Tr}(e^{-\beta\bar H})}
\end{equation}
where $T$ is the time ordering operator, which is bilinear, and is defined in real-space $
  T(c_{j,x,\alpha}(x_0),c_{j',y,\alpha'}^\dagger(y_0))=$
\begin{equation}
  \left\{\begin{array}{>\displaystyle ll}
    c_{j,x,\alpha}(x_0)c_{j',y,\alpha'}^\dagger(y_0)&\mathrm{if\ }x_0<y_0\\
    -c_{j',y,\alpha'}^\dagger(y_0)c_{j,x,\alpha}(x_0)&\mathrm{if\ }x_0\geqslant y_0
    .
  \end{array}\right.
\end{equation}

To compute the 2-point functions we will use the Grassmann integral formalism.
To do so, we add an imaginary time component to all position vectors: we define $\mathbf A_i:=(0,A_i)$.
Given $\mathbf x=(x_0,x)\in \Lambda_j\equiv [0,\beta)\times \mathcal L_j$, we introduce Grassmann variables
 $\psi^\pm_{\mathbf x,a,j}, \psi^\pm_{\mathbf x,b,j}$ and their Fourier transforms
\begin{equation}
  \psi_{j,\mathbf x,\alpha}^\pm
  =\frac1{|\hat\Lambda_j|}\int_{\hat\Lambda_j} d\mathbf k\ e^{\pm i\mathbf k(\mathbf x-\bm\xi)}\hat\psi_{j,\mathbf k,\alpha}^\pm
\end{equation}
with  $|\hat\Lambda_j|=8 \pi^2/3 \sqrt{3}$,
$\bm\xi=(0,\xi)$. The 2-point function can be written as
\begin{equation}
  (S_{j,j'}(\xx,\yy))_{\alpha,\alpha'}:={\int P(d\psi)\ e^{-V(\psi)-M(\psi)}  \psi^-_{j,\xx,\alpha} \psi^+_{j',\yy\,\alpha'}\over 
  \int P(d\psi)\ e^{-V(\psi)-M(\psi)}}
\end{equation}
where $P(d\psi)=P(d\psi_1)P(d\psi_2)$ where $P(d\psi_1)$ is the Grassmann integration with propagator
\begin{equation}
  \hat g_1(\kk)=\begin{pmatrix} -i k_0 &\O(k) \\
  \O^*(k) & -i k_0\end{pmatrix}^{-1}
  .
\end{equation}
and $g_1(\xx,\yy)=\frac t{|\hat{\mathcal L}_1|}\int_{\hat\Lambda_1}d\mathbf k e^{i\kk(\xx-\yy)} \hat g_1(\kk)$
and 
\begin{equation}
 \hat g_1(\kk+K^\pm)\sim \begin{pmatrix} -i k_0  & -v_F(-ik_1\pm k_2) \\
  -v_F (i k_1\pm k_2) & -i k_0\end{pmatrix}^{-1}
  .
\end{equation}
$P(d\psi_2)$ has propagator
\begin{equation}
  g_2(\xx,\yy)={1\over |\hat{\mathcal L}_1|}\int_{\hat\Lambda_2}d\kk e^{i\kk(\xx-\yy)} \hat g_2(\kk)
\end{equation}
with $
  \hat g_2(\kk)=\begin{pmatrix} -i k_0  &\O(R^T k) \\
  \O^*(R^T k) & -i k_0\end{pmatrix}^{-1}
  \equiv
  \hat g_1(R^T \mathbf k)$
which is singular at $K^2_\pm:=R K^1_\pm$ and periodic in $k$ with period $b'_1, b'_2$.
The interaction and counterterm are rewritten formally in terms of Grassmann variables:
\bea\label{112}
V(\psi)=&&
\frac \l{4\pi^2|\hat\Lambda_1|}\sum_{\alpha}\\
&&\cdot 
\left(\sum_{l\in\mathbb Z^2}\int_{\hat\L_1}d \mathbf k \ 
\tau^{(1)}_{l,\alpha}(k+l b)
\hat\psi^+_{1,\mathbf k,\alpha}\hat\psi^-_{2,\mathbf k+l \mathbf b,\alpha}
\right.\nn\\
&&\left.+\sum_{m\in\mathbb Z^2}\int_{\hat\L_2}d\mathbf k\
\tau_{m,\alpha}^{(2)}(k+m b') 
\hat\psi^+_{2,\mathbf k,\alpha}\hat\psi^-_{1,\mathbf k+m \mathbf b',\alpha}\right)
\nn
\eea

\be
M(\psi)=\sum_{\o,\i}
\int   d\kk\ \chi_{\o,i}(k)
(\n_{i,\o}\psi^+_{i,\kk,\a}\psi^-_{i,\kk,\b}+\n^*_{i,\o}\psi^+_{i,\kk,\b}\psi^-_{i,\kk,\a})
\ee

\section{Perturbative expansion and small divisors}
\label{sec:feynman}

\begin{figure}
\hfil\includegraphics[width=8cm]{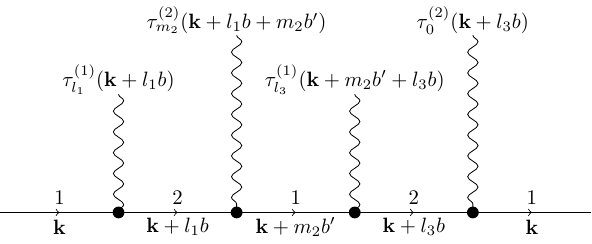}
\caption{\label{fig:feynman} Example of a Feynman diagram for $S_{1,1}(\mathbf k)$ with 4 vertices.}
\end{figure}

To compute the 2-point function $\hat S(\mathbf k)$, we will use Feynman graph expansions. 
The graphs for this model are chain graphs of the form depicted in Figure \ref{fig:feynman}

Let us give the rule first for the graphs with $\nu=0$.
Each line $s$ is associated a layer label $i_s\in\{1,2\}$ and two indices $\alpha_s,\alpha'_s\in\{a,b\}$.
Each vertex has one entering line $s_1$ and an exiting line $s_2$, and we impose that these lines have different indices: $i_{s_1}=3-i_{s_2}$, and the same indices $\alpha_{s_1}=\alpha_{s_2}$.
To each vertex $r$ that has an entering line $s_1$ and exiting line $s_2$, we associate an index, which if $i_{s_1}=1$ we denote by $l_s\in \mathbb Z^2$, and if $i_{s_2}=2$ we denote by $m_s\in \mathbb Z^2$.
\begin{itemize}
  \item
  Each internal line $s$ with layer label $1$ coming from a vertex with index $m_s$ corresponds to the propagator $g_{1;\alpha_s,\alpha'_s}(\mathbf k+m_rb')$ (where $\mathbf k$ is the momentum at which $S$ is being evaluated).
  Each internal line $s$ with layer label $2$ coming from a vertex with index $l_s$ corresponds to the propagator $g_{2;\alpha_s,\alpha'_s}(\mathbf k+l_rb)$.
  \item
  Each external line $s$ corresponds to the propagator $g_{i_s:\alpha_s,\alpha'_s}(\mathbf k)$.
  \item
  Each vertex $r$ that has an entering line $s_1$ and exiting line $s_2$ with
  $i_{s_1}=1$ and $i_{s_2}=2$, if $s_1$ comes from a vertex with index $m_{r-1}$, the vertex $r$ corresponds to an interaction term $\lambda\tau_{l_r,\alpha_{s_2},\alpha'_{s_1}}^{(1)}(\mathbf k+m_{r-1}b'+l_rb)$.
  \item
  Each vertex $r$ that has an entering line $s_1$ and exiting line $s_2$ with
  $i_{s_1}=2$ and $i_{s_2}=1$, if $s_2$ comes from a vertex with index $l_{r-1}$, the vertex $r$ corresponds to an interaction term $\lambda\tau_{m_r,\alpha_{s_2},\alpha'_{s_1}}^{(2)}(\mathbf k+l_{r-1}b+m_rb')$.
\end{itemize}
The value of a graph is obtained by taking the product over the lines of the corresponding propagators, and multiplying them by the product over the vertices of the interaction terms.
The {\it 2-point function} is then obtained by taking a sum over all possible graphs with the correct external labels.
A more explicit expression for the {\it 2-point function} is given in Appendix \ref{app:explicit_feynman}.
The graphs with $\nu\not=0$ are obtained by the above ones in the standard way adding vertices such that the momentum of
the external lines is the same.

The persistence or not of the semimetallic behavior depends on the convergence of the expansion.
One of the more important and difficult reasons for which convergence could, in principle, not occur, is that small divisors could accumulate, as we will now explain.
Note that if $\kk_1$ and $\kk_2$ are 2 neighboring terms their difference is given by $k_1-k_2+m b'+l b$;
moreover the propagator is singular at the Dirac points $K_\o^{i}$. 
Consider therefore  $g_1(\kk) \t^{(1)}_l (k+l b) g_2(\kk+l b)$
and suppose that $k$ is in the first Brillouin zone and is close to 
$K^1_\o$, that is $k=K^1_\o+r_1$ with $r_1=O(\e)$ and $\e$ is a small parameter 
(so that the propagator is $O(\e^{-1})$);
suppose now that $k+l b+mb'$ is also in the first Brillouin zone (by periodicity we can always add $m b'$ to achieve this) 
and close to $K^2_{\o'}$, that is $k+l b+m b'=K^2_\o+r_2$ with $r_2=O(\e)$.
In principle this would produce an $O(\e^{-2} )$ contribution, that is an accumulation of small divisors which
could produce large contributions which can destroy convergence.
Note however that
\be
O(\e)=|r_1|+|r_2|\ge |r_1-r_2|=|l b+m b'-K^2_{\o'}+K^1_{\o}|\label{d1}
\ee
so that this accumulation of small divisors is possible only when the quantity in the r.h.s. is small.
The effect of such terms is rather delicate to be understood.
Indeed 
similar terms in the case of random potential produce a localization phase in 1d,
while an extended phase in higher dimension, while a quasi periodic disorder in 1d produces an extended phase.

\section{Main result}
\label{sec:result}
As we noted above, the accumulation of small divisors can only occur when
$|l b + m b' + K^i_{\o} - K^j_{\o'}|$ is small.
For generic values of $\theta$ (for which $s_\theta$ or $c_\theta$ is irrational), this happens when $l,m$ are large enough, and the basic issue for stability is whether small divisors are balanced by the fact that the interlayer hopping is weak for large values of $l,m$.
We therefore need to quantify the relation between the size of $lb + mb' + K^i_\omega - K^j_{\omega'}$ and that of $l,m$, which we do using number-theoretical considerations.
One such consideration is imposing on the angles a 
{\it Diophantine} condition \pref{cond}. 
We will prove in Lemma \ref{lemm:dioph} below that this condition holds for {\it most} angles, in a sense made precise below.

If this is satisfied, then when $|K^i_\o - K^j_{\o'} + l b + m b'| = O(\epsilon)$, we have
\be
O(\e) \ge C_0 |l|^{-\t}.
\ee
Therefore $|l| \ge \e^{-1/\t}$, and, due to the exponential decay of $\hat \varsigma$ in $\tau^{(1)}$ (see (\ref{interlayer})), we get $|\t^{(1)}_l(k + l b)| \le e^{-\kappa \e^{-1/\t}}$, which compensates the small divisors $\e^{-2}$. This simple argument shows that small divisors due to adjacent propagators do not accumulate.
Of course, this
argument is not sufficient to prove the convergence of the series, and thus the persistence of cones;
it only says that two adjacent propagators cannot be simultaneously small, but it says nothing about non-adjacent ones, which is the generic case.
This requires a Renormalization Group analysis, as we will discuss below. We first present our main result.
\bigskip
\begin{theorem}
For any $\theta$ chosen to satisfy the Diophantine condition (\ref{cond}),
there exists an
$\e_0(C_0)$ such that, for any $|\l| \le \e_0$, there exists $\n_{j,\o}$ 
such that, in the limit $\b\to\infty$ the 2-point function verifies 

\noindent
$
(\hat S_{j,j}(\mathbf k+\mathbf K^j_\o )) =$
\be
\begin{pmatrix}\label{43}
 -i Z_{j,\o} k_0  &  (i v_{j,\o} k_1 - w_{j,\o} \o k_2 )\\
(-i v_{j,\o}^* k_1 - w_{j,\o}^* \o k_2) &   -i Z_{j,\o} k_0
\end{pmatrix}^{-1} (1 + O(|\mathbf k|^{\alpha}))
\ee
with $0 \le \alpha \le 1$, 
$Z_{j,\o}=1 + O(\l)$ real and 
$v_{j,\omega} = 3t/2 + O(\l)$, $w_{j,\omega} = 3t/2 + O(\l)$, 
$\n_{j,\o} = O(\l)$. 
\end{theorem}

The interlayer coupling modifies
the position of the Dirac points; 
we have properly chosen the bare Dirac points $K^j_\pm(\l)$ in the absence of interlayer coupling (given by $\O(K^j_\pm(\l)) + \n_{j,\pm} = 0$)
so that their renormalized physical value is $K^i_\pm$ given by (13).
This is equivalent to say that the position of the Dirac points generically moves depending on the angle, the layer, and the coupling.

The velocities
$w_{j,\o}, v_{j,\o}$  and the wave-function normalization $Z_{j,\o}$
are renormalized in a way generically dependent on the layer and the angle.
Note that a priori several other relevant terms could be present, but they are excluded by symmetry. The singularity
of the 2-point function is given by
$Z_{j,\o}^2 k_0^2 + R(k)$ with $R(k) \sim (|v_{j,\o}|^2 k_1^2 + |w_{j,\o}|^2 k_2^2)$; the singularity of the 2-point function is therefore the same as in the absence of interlayer coupling at weak interaction, proving the stability of the semimetallic phase.

\section{The Diophantine condition} \label{sec:feynman1}

One has, of course, to check the validity of 
\pref{cond}. As the value of the Diophantine constant $C_0$
is expected to depend strongly on the exact value of $\theta$, 
it is convenient to give a generic statement for angles in a certain set. As in the case of disordered systems, we give
a statement in measure (or probability) sense.
We want to estimate the relative 
measure of the set for which \pref{cond} holds. This requires some care as the $K_\pm, b, b'$ are not independent parameters
but given functions of the angles.

As a starting point,
let us consider 
the case $i=j$ and $\omega=\omega'$.
Let
$
M :=
lb + mb'$ and we wish to obtain a lower bound on $|M|$.
To do so, we use a simple inequality: $\forall x,y \in \mathbb R$,
\begin{equation}
  \sqrt{x^2 + y^2} \ge
  \frac{\sqrt3}{2} x - \omega \frac12 y
\end{equation}
and so, since
\begin{equation}
\begin{array}{rl}
M =
\frac{2\pi}{3} (& l_1 + l_2 + m_1 \varphi_1  + m_2 \varphi_2 ,\\&
\sqrt{3} (l_1 - l_2 + m_1 \varphi_3 - m_2 \varphi_4) )
\end{array}
\label{M}
\end{equation}
with $\varphi_1 = c_\theta - s_\theta \sqrt{3}$, $\varphi_2 = c_\theta + s_\theta \sqrt{3}$, $\varphi_3 = c_\theta + s_\theta / \sqrt{3}$, $\varphi_4 = c_\theta - s_\theta / \sqrt{3}$,
we have
\begin{equation}
  |M| \ge
  \frac{2\pi}{\sqrt3} (l_\omega + m \cdot f_\omega)
  \label{Mineq}
\end{equation}
with $l_+ \equiv l_1$ and $l_- \equiv l_2$, and
\begin{equation}
  f_\omega(\theta) := \left( \frac{\varphi_1 - \omega \varphi_3}{2}, \frac{\varphi_2 + \omega \varphi_4}{2} \right) .
\end{equation}
Thus, we wish to impose a condition on $\theta$ such that
$g_{l_\omega,m}(\theta) := | l_\omega + m \cdot f_\omega(\theta)| \geqslant C_1 |m|^{-\tau}$.
The measure of the complement of the 
set where this is true is bounded by
\be
\sum_{m,l_\omega}^* \int_{-C_1 |m|^{-\tau}}^{C_1 |m|^{-\tau}} \frac{1}{g'_{l_\omega,m}} \, dg_{l_\omega,m}
\ee
where $g'_{l_\omega,m}$ is the derivative of $g_{l_\omega,m}$, and $\sum^*_{k,l}$ has the constraint that $\exists \theta \in [\theta_0,\theta_1]$ such that $g_{l_\omega,m}(\theta) \in [-C_1 |k|^{-\tau}, C_1 |k|^{-\tau}]$. Therefore we get the bound, taking into account the sum over $l_\omega$,
\be
\sum^*_{m} 2 C_1 \frac{|m|^{1-\tau}}{\min_{\theta} |m \cdot f_\omega'(\theta)|} \label{kxx}
\ee
Now, in order for this bound to be useful, we need a good bound on $m \cdot f_\omega'$.
Now $m \cdot f_\omega'(\theta)$ can be small, but only if $m$ is large enough.
This suggests we should control it using another Diophantine condition, but we would run into an infinite problem: we would need $m f_\omega''$ to be bounded below, for which we would impose an extra Diophantine condition, which would itself require a Diophantine condition on the third derivatives, et c\ae tera.

We can avoid this problem as follows. Let us fix for definiteness $\theta$ in a small interval around $\bar\theta$.
If
$f_\omega'(\bar\theta) = |f_\omega'(\bar\theta)|(\cos\b, \sin\b)$ is non-vanishing and $\th - \bar\theta$ is small, then
$(m \cdot f_\omega' / |f_\omega'|) \sim |m| \cos(\th_m)$
where $\th_m$ is the angle between $\b$ and $m$.
We can distinguish in the sum (\ref{kxx}) the sum over $m$
such that $|\th_m|, |\th_m - \pi| < \pi/4$ and the complementary set
$|\th_m - \pi/2|, |\th_m - 3\pi/2| < \pi/4$.
In the first sum, $(m \cdot f_\omega'/|f_\omega'|)$ will be greater than $|m|$
up to some constant; we impose a Diophantine condition only for the second term:
for $|\th_m - \pi/2|, |\th_m - 3\pi/2| < \pi/4$ we assume that 
$|m \cdot f_\omega'| \le C_1 |m|^{-\t}$.
Again we will end up with a condition like (\ref{kxx})
involving 
$m \cdot (f_\omega'/|f_\omega'|)'$ which in principle could be arbitrarily small.
However $(f'/|f'|)'$
is orthogonal to $f_\omega'/|f_\omega'|$, hence 
$m \cdot (f_\omega'/|f_\omega'|)' \sim |m| \cos(\th_m + \pi/2)$ 
which, in this region, is greater than $|m|$. Note that the derivatives $f'(\bar\theta)$ must be finite.

Thus, we construct a large-measure set of $\theta$'s that satisfy $|l_\omega + m \cdot f_\omega(\theta)| \geqslant C_1 |m|^{-\tau}$.
More exactly,
in appendix \ref{app:dioph}, we prove the following lemma (this proof is computer-assisted).

\begin{lemma}\label{lemm:dioph}
For any $\theta \in [0,2\pi) \setminus (\mathcal B_1 \cup \mathcal B_2)$ where $\mathcal B_1 \cup \mathcal B_2$ is a finite set of special angles (see (\ref{B1}) and (\ref{B2}) for an explicit expression), and any $\delta \theta$ small enough, for almost every angle in $[\theta - \delta \theta, \theta + \delta \theta]$, there exists $C_0$ such that, for all $\omega,\omega',i,j$, and for all $l,m \in \mathbb Z^2 \setminus \{0\}$,
\be
|K^i_\o - K^j_{\o'} + l b + m b'| \ge \frac{C_0}{|m|^\t}
,\quad
|K^i_\o - K^j_{\o'} + l b + m b'| \ge \frac{C_0}{|l|^\t}
.
\ee
\end{lemma}

\section{The renormalized expansion}\label{sec:renormalized}

\subsection{Multiscale decomposition}
\label{sec:multiscale}

We introduce smooth cut-off functions: for $i=1,2$, $\o=\pm$, $h\in\{-\infty,\cdots,0\}$, let $\chi_{h,i,\o}(\kk)$ be a smooth function that vanishes outside the region $||\kk-\mathbf K^i_\omega|| \le \g^{h-1}$ and that is equal to 1 for $||\kk-\mathbf K^i_\omega||\ge \g^{h-2}$.
The constant $\g>1$ will be chosen below to be large enough. Note that, in this way, the supports of $\chi_{0,i,+}$ and $\chi_{0,i,-}$ do not overlap.
We define $\hat g_{i,\o}^{(\le 0)}(\kk)=\chi_{0,i,\o}(\kk)\hat g_i(\kk)$
and 
\be \hat g_i(\kk)= g_i^{(1)}(\kk)+\sum_{\o=\pm} \hat g^{(\le 0)}_{i,\o} (\kk)
\ee
with $\hat g^{(1)}(\kk)=(1-\sum_\o \chi_{0,i,\o}(k))\hat g_i(\kk)$; this induces the Grassmann variable decomposition $\hat\psi_{i,\kk,\alpha}=\hat\psi_{i,\kk,\alpha}^{(1)}+\sum_{\o=\pm} \hat\psi^{(\le 0)}_{i,\kk,\alpha,\o}$
with propagators given by  $\hat g^{(1)}_i$ and $\hat g^{(\le 0)}_{i,\o}$
respectively. Note that $\hat\psi^{(1)}$ correspond to fermions with momenta far from the Fermi points, while  $\hat\psi^{(\le 0)}$ with momenta around $K_{\pm}$.

We further decompose 
\be
\hat g_{i,\o}^{(\le 0)} (\kk)=\sum_{h=-\io}^0 
\hat g^{(h)}_{i,\o}( \kk)
\ee
where $\hat g_{i,\o}^{(h)}(\kk):=f_{h,i,\o}(\kk)\hat g_{i,\o}^{(\le 0)}$ in which $f_{h,i,\o}:=\chi_{h,i,\o}-\chi_{h-1,i,\o}$ is a smooth cutoff function supported in $ \g^{h-3} \le |\kk-\KK_\o^{i}|\le  \g^{h-1}$ such that $\sum_{h=-\io}^0f_{h,i,\o}=\chi_{0,i,\o}$. 
The integration is done recursively in the following way: assume that we have integrated the fields $\psi^{(1)},..,\psi^{(h-1)}$ obtaining
\be
e^{W(\phi)}=\int\bar P(d\psi^{(\le h)}) e^{V^{(h)}(\psi,\phi)}  
\ee
where $\bar P(d\psi^{(\le h)})$ is Gaussian integration with propagator $\bar g_{i,\omega}^{(\le h)}$ which will be defined inductively in (\ref{prop_ind}),
and
\bea
&&V^{(h)}(\psi,0)=
\\
&&\sum_{i,\o,\o',l,\alpha,\alpha'} \int_{\hat\L_i} d\kk W^{(h,\omega,\omega')}_{i,2,l,\alpha,\alpha'}(\kk) 
\psi_{i,\kk,\alpha,\o}^+\psi_{2,\kk+lb,\alpha',\o'}^-+\nn\\
&&\sum_{i,\o,\o',m,\alpha,\alpha'} \int_{\hat\L_i}  d\kk W^{(h,\omega,\omega')}_{i,1,m,\alpha,\alpha'}(\kk) 
\psi_{i,\kk,\alpha,\o}^+\psi_{1,\kk+mb',\alpha',\o'}^-
.\nn
\label{eff}
\eea

According to the RG procedure, we renormalize the relevant and marginal terms; we will see below that the term
with $l$ or $m$ non zero are actually irrelevant, due to improvements in the estimates due to the Diophantine condition. 
We therefore define a localization operation in the following way
\bea
&&\LL W^{(h,\omega,\omega')}_{i,j,l}(\kk)=\d_ {\o,\o'}\d_{i,j} \d_ {l,0}[W^{(h,\omega,\omega)}_{i,i,0}(0,K^i_\o)+\nn\\
&&k_0 \partial_0 W^{(h,\omega,\omega)}_{i,i,0}(0,K^i_\o)+
(k-K^i_\o)\partial  W^{(h,\omega,\omega)}_{i,i,0}(0,K^i_\o)
.
\label{loc}
\eea
The terms for which $\LL=0$ are called {\it non resonant} terms and the ones for which $\LL\not=0$ {\it resonant} terms. The terms containing derivatives
are marginal ones and produce wave function or velocities renormalizations, while the terms without derivatives are the relevant terms. 
The action of $\mathcal L$ on the effective potential $V^{(h)}$ is
\begin{equation}
  \LL V^{(h)}=\LL_1 V^{(h)}+\LL_2 V^{(h)}
\end{equation}
with
\begin{equation}
\LL_1 V^{(h)}:=\sum_{i,\omega,\a,\a'}\int_{\hat \Lambda_i} d\kk \g^h\n_{h,\omega,\a,\a',i}
\psi^+_{\kk,\o,i,\a}\psi^-_{\kk,\o,i,\a'}
\end{equation}
and
\begin{equation}
\LL_2 V^{(h)}:=\sum_{i,j,\omega,\a,\a'}\int_{\hat \Lambda_i} d\kk
z_{h,\o,\a,\a',i,j} (\mathbf k-\mathbf K^i_\omega)_j\psi^+_{\kk,\o,i,\a} \psi^-_{\kk,\o,i,\a'}
\end{equation}
with
\begin{equation}\label{ai}
  \nu_{h,\omega,\alpha,\alpha',i}:=\gamma^{-h} W_{i,i,0,\alpha,\alpha'}^{h,\omega,\omega}(0,K^i_\omega)
\end{equation}
\begin{equation}
  z_{h,\omega,\alpha,\alpha',i,j}:=-\partial_{\mathbf k_j}W_{i,i,0,\alpha,\alpha'}^{h,\omega,\omega}(0,K^i_\omega)
  .
\end{equation}

The form of the resonant terms is severely constrained by symmetries: as is proved in Appendix \ref{sec:symm},
\bea
&&\n_{h,\o,a,a,i}=\n_{h,\o,b,b,i}=0\quad \n_{h,\o,a,b,i}=\n^*_{h,\o,b,a,i}\nn\\
&&z_{h,\o,b,a,i,1}=z_{h,\o,a,b,i,1}^*\quad
z_{h,\o,b,a,i,2}=z_{h,\o,a,b,i,2}^*
\\
&&z_{h,\o,a,a,i,1}=z_{h,\o,b,b,i,1}=
z_{h,\o,a,a,i,2}=z_{h,\o,b,b,i,2}=0
\nn
\\
&&z_{h,\o,b,a,i,0}=z_{h,\o,a,b,i,0}=0
\quad
z_{h,\o,a,a,i,0}=z_{h,\o,b,b,i,0}\in i \mathbb R
\nn
\eea

The contributions from $\mathcal L_2 V$ are marginal, and are absorbed into the propagator at every step of the integration:
\begin{equation}
  \begin{array}{>\displaystyle l}
  \bar g_{i,\omega}^{(\le h)}(\mathbf k)
  :=
  \chi_{h,i,\omega}(\mathbf k)
  \cdot\\\hfill\cdot
  \left((\bar g_{i,\omega}^{(\le h+1)}(\mathbf k))^{-1}
  -\sum_j z_{h,\omega,\cdot,\cdot,i,j}(\mathbf k-\mathbf K^i_\omega)_j\right)^{-1}
  \end{array}
  \label{prop_ind}
\end{equation}
Thus,
\begin{equation}
  \begin{array}{>\displaystyle l}
\bar g_{i,\o}^{(\le h)}(\kk+\mathbf K^i_\o)=
\chi_{h,i,\o}(\kk)(1+O(\kk))
\cdot\\\hfill\cdot
\begin{pmatrix}
 -i Z_{i,\o,h} k_0  &  (i v_{i,\o,h} k_1- w_{i,\o,h} \o k_2)\\
(-i v^*_{i,\o,h} k_1- w^*_{i,\o,h} \o k_2) &   -i Z_{i,\o} k_0
\end{pmatrix}
^{-1}\label{prop}
\end{array}
\end{equation}
with
\bea
  &&Z_{i,\omega,h}=Z_{i,\omega,h+1}-iz_{h,\o,a,a,i,0}
  \nn\\
  &&v_{i,\omega,h}= v_{i,\omega,h+1}+i z_{h,\o,a,b,i,1}
  \nn\\
  &&w_{i,\omega,h}= w_{i,\omega,h+1}+ \omega z_{h,\o,a,b,i,2}
  .
\eea

After absorbing $\mathcal L_2V^{(h)}$ into the propagator, we are left with integrating $\LL_1 V^{(h)}$ and
\begin{equation}
  \RR V^{(h)}:=(1-\mathcal L)V^{(h)}
\end{equation}
so
\be
e^{W}=\int \bar P(d\psi^{(\le h)}) e^{\LL_1 V^{(h)}(\psi)+\RR V^{(h)}(\psi)}  
.
\label{renormalized_expansion}
\ee

\subsection{Feynman rules for the renormalized expansion} \label{sec:renormfeyn}

The renormalized expansion described above has a graphical representation that is similar to the Feynman diagram expansion from Section \ref{sec:feynman}.
There are two main differences: first, there are two different types of vertices: ``\emph{$\tau$-vertices}'', coming from $\mathcal R V^{(h)}$ in (\ref{renormalized_expansion}), and ``\emph{$\nu$-vertices}'', coming from $\mathcal L_1 V^{(h)}$.
Second, every line has a {\it scale label} $h$, corresponding to a propagator on scale $h$:
\begin{equation}
  \bar g_{i,\omega}^{(h)}(\mathbf k):=f_{h,i,\omega}(\mathbf k)\bar g^{(\le h)}_{i,\omega}(\mathbf k)
  .
\end{equation}

\begin{figure}
\hfil\includegraphics[width=8cm]{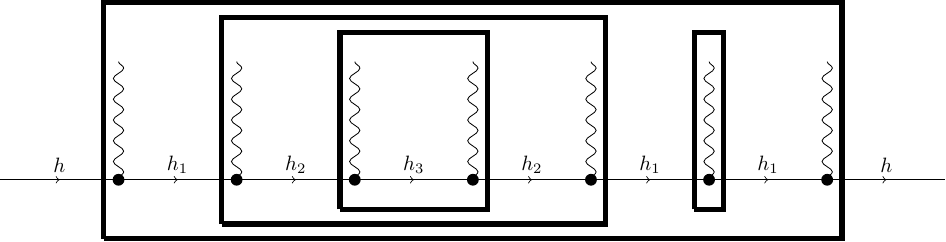}

\caption{\label{fig:clusters} An example of graph of order $\l^7$ with the associated clusters, denoted by thick rectangles. In this example, $h<h_1<h_2<h_3$.}
\end{figure}

The scale labels induce an important structure: given a diagram, we group vertices together into nested \emph{clusters}, which are connected subgraphs in which the scales of the lines leaving the cluster are all smaller than the scales of the lines inside the cluster, see Figure \ref{fig:clusters}.
A cluster that is such that $\mathcal L$ applied to the cluster yields $0$ is called {\it non-resonant}, otherwise it is called {\it resonant}.
In other words, the action of $\mathcal R=1-\mathcal L$ is trivial on non-resonant clusters, and non-trivial on resonant ones.

Some clusters are single vertices (either $\nu$ or $\tau$) and are called \emph{trivial clusters}.
The clusters that contain internal lines are called \emph{non-trivial clusters}.
As per the construction above, if $h^{ext}_T$ is the largest of the scales of the external lines of a non-trivial cluster $T$, all its internal lines have a scale $h>h^{ext}_T$; $h_T$ is the largest scale of the propagators internal to the cluster $T$.
A non-trivial cluster $T$ contains sub-clusters $\tilde{T}\subset T$.
We call a cluster $\tilde T\subset T$ a \emph{maximal cluster}
if there is no other cluster $\bar T$ such that  $\tilde T\subset \bar T\subset T$. See e.g. \cite{M10}, \cite{M11}, \cite{M12} for more details.

For each cluster, there are two external lines that connect the cluster to other ones.
In a non-resonant cluster $T$ with external lines of type $i_1=i_2=1$ and momenta
$k_1, k_2$ with $k_1$ in the first Brillouin zone
and $k_2=k_1+\hat m_T b+l b$ where $l b$ is chosen so that $k_2$ is in the first Brillouin zone,
if $A_0,..,A_N$
are the momenta associated to the $\t$ vertices contained in $T$ (the $\nu$ vertices do not change momentum)
one has $A_0=k_1+ l_0 b$, $A_1=k_1+l_0 b+m_1 b'$, $A_2=k_1+m_1 b'+l_2 b$, $A_3=k_1+m_3 b'+l_2 b$,..., $A_{N}=k_1+m_{N} b'+l_{N-1} b$ and 
$m_{N}=\hat m_T$
with $N$ odd.
In the same way if the non-resonant cluster $T$ has one external line of type $i_1=1$ and momentum $k_1$, and one of type $i_2=2$ and momentum
$k_2$; assume that $k_1$ is in the first Brillouin zone
and  $k_2=k_1+\hat l_T b+m b'$,
where $m b'$ is chosen such that $k_2$ is in the first Brillouin zone.
Now with $N$ even 
the momenta associated to the $\t$ vertices in $T$ are
$A_0=k_1+ l_0 b$, $A_1=k_1+l_0 b+m_1 b'$, $A_2=k_1+m_1 b'+l_2 b$, $A_3=k_1+m_3 b'+l_2 b$,...$A_{N-1}=k+m_{N-2} b'+l_{N-1}b$,
$l_{N-1}=\hat l_T$.

The value associated to a graph $\G$ is denoted by
$W_\G(\kk)$ and is given by the product of the propagators and $\n,\t$ factors associated to the vertices, with the $\RR$ operation acting on each 
non-resonant cluster. The effect of the $\RR$ operation on the non-resonant clusters can be written as
\be
\RR \hat W^h(\kk+\mathbf K^i_\o)=k^2\int_0^1 \partial^2 \hat W(t \kk)\ee

\subsection{Convergence of the series and persistence of the cones}

In this section we want to prove the convergence of the series, extending methods previously used 
\cite{M10}, \cite{M11}, \cite{M12} for fermions in a quasi-periodic potential.
In order to bound the contribution of a Feynman graph we note that $|\bar g^{(h)}(k)|\le C \g^{-h}$ (by (\ref{prop})); therefore the product of propagators in the Feynman graph is bounded by
\be
\prod_{T\mathrm{\ n.t.}} \g^{-h_T (M_T+R_T-1)}\ee
where $M_T$ is the number of maximal non-resonant clusters contained in $T$, $R_T$ is the number of maximal resonant clusters contained in $T$, and $\prod_{T \mathrm{\ n.t.}}$ is a product over non-trivial clusters.
Note that the trivial clusters do not contribute, as they contain no internal lines.
The effect of the $\RR:= \mathds 1-\mathcal L$ (recall (\ref{loc})) operation on the non-trivial resonant clusters 
produces an extra $\g^{2 (h_{T}^{\text{ext}}-h_T)}$ (the gain $h_T^{\mathrm{ext}}$ comes from the extra $(k-K^i_\omega)^2$ terms (which are on the scale of the external lines) and the $h_T$ loss from the extra second derivative (which are on the scale of the internal lines)).
Therefore, we get an extra factor:
\be
\prod_{T\ \mathrm{res\ n.t.}} \gamma^{2(h_{T}^{\text{ext}}-h_{T})}
\ee
where $\prod_{T\ \mathrm{res\ n.t.}}$ is the product over the non-trivial resonant clusters.
In addition, each maximal resonant cluster corresponds to a weight $\gamma^h\nu_h$, so, if $|\nu_h| \leqslant C |\lambda|$, we get an extra factor:
\begin{equation}
  \prod_{T \mathrm{\ n.t.}}\gamma ^{h_T M_T^\nu} C |\lambda|
\end{equation}
where $M_T^\nu$ is the number of maximal resonant trivial clusters in $T$.
Thus, we get the following bound for the sum over all the labels of a renormalized graph with $q$ vertices:
\bea
&&\sum^*_{\underline h,\underline l, \underline m} [L(\underline l, \underline m)] 
|\l|^q C^q
\left( \prod_{T\, \text{n.t.}} \g^{-h_T (M_T+R_T-1)}
\right) \nn\\
&&\Bigg(\prod_{T\ \mathrm{res\ n.t.}} \gamma^{2(h_{T}^{\text{ext}}-h_{T})}\Bigg)
\prod_{T  \, \text{n.t.}} \gamma^{h_{ T} M^\n_{T}}
\label{lap} 
\eea
where $L(\underline l, \underline m)$ is the norm of the product of the $\t$ functions:
\begin{equation}
  L(\underline l,\underline m):=\prod_i| \t(A_i)|
\end{equation}
where $A_i$ are the momenta, which depend on the structure of the graph, and $\sum^*_{\underline h,\underline l, \underline m}$ is the sum over the scales and momenta associated to the vertices, as explained in Section \ref{sec:renormfeyn}; 
these sums are not independent, as they are constrained by the compact support properties of the propagators and of the Diophantine condition.
Now
        defining $\mathds 1_{\Gamma\,\mathrm{res}}$ is equal to 1 if the maximal cluster $T_m$ is resonant and 0 otherwise, we have
\bea
&&\prod_{T\, \text{n.t.}}  \g^{-h_T R_T}  \prod_{T\ \mathrm{res\ n.t.}, T\not=T_m} 
\gamma^{h_{T}^{\text{ext}}} \prod_{T\, \text{n.t.}}
\gamma^{h_{ T} M^\n_{T}}\le 1\nn\\\
&&\prod_{T\, \text{n.t.}}  \g^{h_T}
\prod_{T\ \mathrm{res\ n.t.}, T\not= T_m} 
\gamma^{-h_{T}}
\le \gamma^{h_{T_m} \mathds 1_{\Gamma\,\mathrm{res}}}
\eea 
and then using that 
if $T_m$ is resonant and $\LL$ is applied then $h_{T_m}=h+1$ we get
\be
\prod_{T\, \text{n.t.}}  \g^{-h_T (R_T-1)} \gamma^{h_{ T} M^\n_{T}}\prod_{T\ \mathrm{res\ n.t.}} 
\gamma^{(h_{T}^{\text{ext}}-h_{T})}
\le \gamma^{(h+1)\mathds 1_{\Gamma\,\mathrm{res}}}.\ee 

Therefore, \pref{lap} is bounded by 
\bea
&&\sum^*_{\underline h,\underline l, \underline m} [L(\underline l, \underline m)] 
|\l|^q C^q
\gamma^{h \mathds 1_{\Gamma\,\mathrm{res}}}
\nn\\&&
\left( \prod_{T\, \text{n.t.}} \g^{-h_T M_T}
\right)
\Bigg(\prod_{T\ \mathrm{res\ n.t.}} \gamma^{(h_{T}^{\text{ext}}-h_{T})}\Bigg)
.
\label{lap1} 
\eea
As the graphs are chains, there is no problematic combinatorial factor; the main issue
is the sum over $\underline h$; if we neglect the constraint in 
$\sum^*_{\underline h,\underline l, \underline m}$ then we will get a factor  $
\prod_{T\, \text{n.t.}} 
 (\sum_{h_T=-\io}^0 \g^{-h_T})$ which is divergent.

However, we have still not taken advantage of the Diophantine condition. In order to do that we note that 
$
|\t^{(i)}(k)|\le C e^{-\kappa |k|}
$
from the exponential decay or $\hat \varsigma(k)$, see (\ref{interlayer}), (\ref{tau1})-(\ref{tau2}); we can write 
\be
e^{-\kappa/2 |k|}=\prod_{h=-\io}^0 e^{-\kappa 2^h |k|}
\ee Consider the product of $\t$ factors:
\be
L(\underline l,\underline m)\equiv\prod_i| \t(A_i)|\le \prod_i e^{-\kappa |A_i|/2}
\prod_{h=-\infty}^1 e^{-\kappa 2^{h} |A_i|}\ee
which we estimate as
\be
L(\underline l,\underline m)\le \prod_i e^{-\kappa |A_i|/2}
\prod_{T\ \mathrm{nonres}\ni i} e^{-\kappa 2^{h_T} |A_i|}\ee
where in the last product $i$ is the label of the points $\t$ contained in $T$.
We then exchange the products:
\be
L(\underline l,\underline m)\le \left(\prod_i e^{-\kappa |A_i|/2}\right)
\prod_{T\ \mathrm{nonres}} \prod_{i\in T}e^{-\kappa 2^{h_T} |A_i|}\ee
Let us consider a non resonant cluster with external lines of type $1$;
we have
\be
|A_0|+|A_1|+\ldots\ge  |A_0-A_1+A_2-A_3\ldots|= |\hat m_T b'|\ee
(see Section \ref{sec:renormfeyn}) so that
\be
\prod_{i\in T}e^{-\kappa 2^{h_T} |A_i|}
\le e^{-\kappa 2^{h_T} |\hat m_T b'|}
\ee
A similar analysis holds for $i_1=i_2=2$ with $\hat l\not=0$.
In the same way if the cluster $T$ has $i_1=1$ and $i_2=2$ and 
$k_1, k_2$ are the momenta of the external lines
using that $|A_0|+|A_1|+\ldots$
\be\ge |A_0-A_1+A_2-A_3\ldots|\ge |k_1+\hat l b|\ge |\hat lb|-\frac{4\pi}3\ee
(where we used that $|k_1|\le\frac{4\pi}3$) as $k_1$ is in the first Brillouin zone
\be
\prod_{i\in T}e^{-\kappa 2^{h_T} |A_i|}\le
e^{-\kappa 2^{h_T} (|\hat l_T b|-\frac{4\pi}3)}
.
\ee
In addition, the Diophantine condition imposes a constraint between the momenta
of the external lines of a cluster $T$ and the $l,m$ labels associated to its internal vertices.
Consider a non resonant cluster with the following indices:
\begin{enumerate}
\item If $i_1=i_2=1$ and 
$k_1, k_2$ are the momenta of the external lines; assume that $k_1$
is in the first Brillouin zone
and $k_1=:\bar k_1+K^1_{\o_1}$. Moreover $k_2:=k_1+\hat m_T b'+l b=:\bar k_2+K^1_{\o_2}$,
with $l$ chosen such that $k_2$ is in the first Brillouin zone; then, if $\hat m_T\not=0$, by (\ref{cond}),
\bea
&&2 \g^{h^{ext}_T}\ge |\bar k_1|+|\bar k_2|\ge |\bar  k_1-\bar k_2|=\nn\\
&&|
K^1_{\o_1}-K^1_{\o_2}+\hat m_T b'+l b|\ge {C_0\over |\hat m_T|^\t}\eea
so that, if $\hat m_T\not=0$
\be
|\hat m_T|\ge ({\textstyle\frac12}C_0 \g^{-h^{\mathrm{ext}}_T})^{\frac1\tau}\label{cond1}
\ee
and so
\begin{equation}
  |\hat m_Tb'|\ge c_1\g^{-h^{\mathrm{ext}}_T/\tau}
\end{equation}
for some constant $c_1>0$.

On the other hand if $\hat m_T=0$ but $\o_1\not=\o_2$ then $l=0$ so $\g^{h^{\mathrm{ext}}_T}\ge\frac12|K^1_{\omega_1}-K^1_{\omega_2}|=\frac{2\pi}{3\sqrt3}$ (recalling (\ref{pF})).
Thus, this eventuality does not occur provided $\gamma$ is large enough.

\item
In the case $i_1=1$ and $i_2=2$ and 
$k_1, k_2$ are the momenta of the external lines; assume that $k_1$  is in the first Brillouin zone
and $k_1=:\bar k_1+K^1_{\o_1}$. Moreover $k_2:=k_1+\bar l b+m b'=:\bar k_2+K^2_{\o_2}$,
with $m$ chosen in such a way that $k_2$ is in the first Brillouin zone; then, if $\bar l\not=0$, by (\ref{cond}),
\bea
&&2 \g^{h^{ext}_T}\ge |\bar k_1|+|\bar k_2|\ge |\bar  k_1-\bar k_2|=\\
&&|K^1_{\o_1}-K^2_{\o_2}
+\hat l_T b+m b'|\ge {C_0\over |\hat l_T|^\t}\nn\eea
so that
\be|\hat l_T|\ge ({\textstyle\frac12}C_0 \g^{-h^{\mathrm{ext}}_T})^{\frac1\tau}\label{cond2}
\ee
and so
\begin{equation}
  |\hat l_Tb|\ge c_1\g^{-h^{\mathrm{ext}}_T/\tau}
\end{equation}
 If $\hat l_T=0$ then $2 \g^{h^{\mathrm{ext}}_T}\ge O(\th)$ for $\o_1=\o_2$ and $2 \g^{h^{\mathrm{ext}}_T}\ge O(1)$ for $\o_1=-\o_2$.
Thus, provided $\gamma\gg \theta^{-1}$, these eventualities do not present themselves provided $\gamma$ is large enough.

\item
A similar analysis holds for $i_1=2, i_1=1$, and $i_1=i_2=2$.
\end{enumerate}

Thus,
\begin{equation}
  \prod_{i\in T}e^{-\kappa 2^{h_T}|A_i|}\le e^{-\kappa 2^{h_T}(c_1 \gamma^{-h_T^{\mathrm{ext}}/\tau}-\frac{4\pi}3)}
\end{equation}
which, provided $\gamma$ is large enough, yields
\begin{equation}
  \prod_{i\in T}e^{-\kappa 2^{h_T}|A_i|}\le e^{-c_2\gamma^{-h_T^{\mathrm{ext}}/\tau}}
\end{equation}
for some constant $c_2$; note that such constant is proportional, up to some numerical constant, to $C_0^{\t}$, as follows from (\ref{cond1}). 
Therefore,
\be
L(\underline l,\underline m)\le e^{-c_2 \gamma^{-h/\tau}\mathds 1_{\Gamma\,\mathrm{nonres}}}
\prod_i e^{-\kappa |A_i|/2}
\prod\limits_{T\ \mathrm{n.t.}}
e^{-c_2 {M}_{T} \gamma^{-h_{T}/\tau}}\ee
where $\mathds 1_{\G\,\mathrm{nonres}}$ is equal to 1 if the maximal cluster is non resonant and $0$ otherwise.
Note that, provided $\gamma$ is chosen to be large enough, $e^{-c_2 \gamma^{-h/\tau}\mathds 1_{\Gamma\,\mathrm{nonres}}}\le \gamma^{3h \mathds 1_{\Gamma\,\mathrm{nonres}}}$, so
\be
L(\underline l,\underline m)\le \gamma^{3h\mathds 1_{\Gamma\,\mathrm{nonres}}}
\prod_i e^{-\kappa |A_i|/2}
\prod\limits_{T\ \mathrm{n.t.}}
e^{-c_2 {M}_{T} \gamma^{-h_{T}/\tau}}.\ee

Furthermore, using the bound $e^{-\a x}\le ({\beta\over \a})^\beta e^{-\beta}x^{-\beta}$ with $\beta= 3\tau M_T$, we find
\begin{equation}
  e^{-c_2M_T \gamma^{-h_T/\tau}} \leq (\frac{c_2 e^1}{3\tau})^{-3\tau M_T} \gamma^{3M_T h_T}
  .
  \label{exp3}
\end{equation}
In addition, $\sum_{T\,\text{n.t.}} M_T \leq q$, since the clusters are nested in each other and for two clusters to be different they must differ by at least one vertex. 

Now, let us introduce $M_T^\tau$ as the number of maximal non-resonant trivial clusters (i.e. maximal $\tau$-vertices) contained in $T$, and use the trivial bound $3M_T \le 2M_T+M_T^\tau$ along with (\ref{exp3}) to obtain
\begin{equation}
\prod_{T\ \mathrm{n.t.}}
e^{-c_2 M_{T} \gamma^{-h_{T}}/\tau} \le  C_3^q
.\prod_{T\ \mathrm{n.t.}} \gamma^{h_{T} (2M_{T}+M_T^\tau)}
\label{asxq}
\end{equation}
Thus, plugging this into (\ref{lap1}), we find
\bea
&&\g^h \sum^*_{\underline h,\atop \underline l, \underline m}
[L]^{1\over 2} |\l|^q (CC_3)^q
\gamma^{h\mathds 1_{\Gamma\,\mathrm{res}}}
\gamma^{3h\mathds 1_{\Gamma\,\mathrm{nonres}}}
\nn\\&&
[\prod_{T\ \mathrm{res}}
\gamma^{(h_{T}^{\text{ext}}-h_{T})}]
\prod_{T\, \text{n.t.}} \g^{h_T (2M_T+M^\tau_T)}
.
\eea
In addition,
\begin{equation}
  \prod_{T\ \mathrm{n.t.}} \gamma^{2h_T M_T}
  =
  \gamma^{-2h \mathds 1_{\Gamma\,\mathrm{nonres}}}\prod_{T\ \mathrm{nonres}}\gamma^{2h_T^{\mathrm{ext}}}
\end{equation}

\be
\g^h \sum^*_{\underline h,\atop \underline l, \underline m} [L]^{1\over 2} |\l|^q (CC_3)^q
[\prod_{T\, \text{n.t.}}
\gamma^{(h_{T}^{\text{ext}}-h_{T})}]
\prod_{T\,\text{n.t.}} \gamma^{h_{T} M^\tau_{T}} \label{ssa}
\ee
The crucial point is that the sum over the scales $h$ can be performed
summing over all the differences, taking into account that the scale $h$ is fixed.
Finally the sum over the $l,m$ is done using the factor $[L(\underline l, \underline m)]^{1\over 2}$.
(The gain term $\g^{h_T M_T^\tau}$ is dropped, as it does not lead to any significant gain.)
In conclusion the bound on a graph with $q$ vertices is $C_4^q\g^h |\l|^q$ assuming that $|\n_h|,|Z_h-1|,|v_h-1|,|w_h-1| \le C |\l|$.
The constant $C_4$ is proportional to $C_0^{4 \t}$ as one can check collecting the estimates above.

\subsection{Beta function and imaginary time Green functions}

We are left with checking our assumption on $Z_h, \nu_h, w_h,v_h$.
We know that 
 $v_{i,\omega,h}= v_{i,\omega,h+1}-i z_{h,\o,a,b,i,1}$
with $z_{h,\o,a,b,i,1}$ expressed by the sum of renormalized Feynman graphs $\G$ such that the maximal scale of the clusters is $h+1$, an extra derivative is applied (which costs a factor $\gamma^{-h}$)
and the momenta of the external lines is fixed equal to $K_\o$.
Moreover by the compact support of the propagator there is at least a $\t$ vertex, as the $k=0$ value of a graph with only $\n$ vertices is zero; therefore the
analogue of (\ref{ssa}) becomes
\be
\sum^*_{\underline h,\atop \underline l, \underline m} [L]^{1\over 2} |\l|^q (CC_3)^q    [\prod_{T\, \text{n.t.}}
\gamma^{(h_{T}^{\text{ext}}-h_{T})}]
\gamma^{2h_{T^*}}
\label{weightzz}
\ee
where $T^*$ is the non trivial cluster containing a $\t$ vertex whose scale is the largest possible (we now use the gain $\g^{h_T M_T^\tau}$  dropped in the bound \pref{ssa}).
In addition, summing the differences $h_T^{\mathrm{ext}}-h_T$ along a sequence of clusters that goes from $h$ to $h_{T^*}$ and discarding the others, we bound
\begin{equation}
  \sum_{T}(h_T^{\mathrm{ext}}-h_T) \leqslant h-h_{T^*}
\end{equation}
and so (\ref{weightzz}) is bounded by
\be \label{weightzz1}\g^{h\over 2}
\sum^*_{\underline h,\atop \underline l, \underline m} [L]^{1\over 2} |\l|^q (CC_3)^q    \prod_{T\, \text{n.t.}}
\gamma^{ {1\over 2}(h_{T}^{\text{ext}}-h_{T})}r
.
\ee
Estimating the sum as above, we find that 
\be\label{app}
|z_{h,\o,a,b,i,1}|\le C_5 \l  \g^{h\over 2}\quad 
v_{i,\omega,h}= v_{i,\omega,0}-i \sum_{h'} z_{h',\o,a,b,i,1}
\ee
hence  $v_{i,\omega,h}= v_{i,\omega,0}+O(\l)$; moreover the limiting value is reached exponentially fast 
$v_{i,\omega,h}=v_{i,\omega,-\infty}+O(\l \g^{h/2})$. At the end $v_{i,\omega,-\io}$
is expressed by a sum over renormalized graphs; the lowest order contribution, after summing over scales,
contain the derivative of (considering for definiteness the external lines with layer label $1$)
a term proportional to the sum of $\l^2\tau_{m}^{(1)}(K^1_\o+l b)   g_2(K^1_\o+l b+m b')\tau_{m,1}^{(2)}(K^1_\o+l b)$ 
where $m b'$ is chosen so that $K^1_\o+l b+m b'$ is in the first Brilloin zone. 
In the case of small angles
we can separate in the sum the contribution
a) $m_1=m_2=0$,
 $l_1=l_2=0$; b) $m_1=-l_1=1$ and $m_2=l_2=0$; c)$m_2=-l_2=1$ and $m_1=l_1=0$.
In the above three cases one has $\partial g_2(K+m b'+l b)=O(\th^{-2})$; actually one can decompose the propagator
in a dominant part, with a result coinciding with the effective continuum model 
and whose contribution is a constant times $\l^2/\th^2$, plus a correction which is 
bounded by $C \l^2$. The contribution from the other moments is bounded as 
$\l^2 \sum_l  C C_0 |l|^\t e^{-\x |l|}$ and the higher orders are $O(\l^3)$ by \pref{ssa}.
A similar argument holds for $Z_h, w_h$. 

It remains to discuss the flow of $\n_h$; we can write, see  
\pref{ai},
\be W_{i,i,0,\alpha,\alpha'}^{h,\omega,\omega}(0,K^i_\omega)
=\g^{h+1}\n_{h+1}+ \tilde W_{i,i,0,\alpha,\alpha'}^{h,\omega,\omega}(0,K^i_\omega)\ee where $\tilde W$ is given by the sum of the terms with a number of vertices greater or equal to $2$; therefore 
\be
\nu_{h,\omega,\alpha,\alpha',i}=\g \nu_{h+1,\omega,\alpha,\alpha',i}+\b^h_\n
\label{appo2}
\ee
with $\b^h_\n=\g^{-h}
\tilde W_{i,i,0,\alpha,\alpha'}^{h,\omega,\omega}(0,K^i_\omega)$ 
is given by the sum with $q\ge 2$ of terms bounded by \pref{weightzz1}. 
We have to prove that it is possible to choose the counterterms $\n_{\omega,\alpha,\alpha',i}$ so that  $\n_{h,\omega,\alpha,\alpha',i}$ is bounded by $C \l$ for any scale $h$. Indeed from \pref{appo2} we get, $h\le -1$ 
\be
\nu_{h,\omega,\alpha,\alpha',i}=\g^{-h}(\nu_{\omega,\alpha,\alpha',i}+
\sum_{i=h}^{-1} \g^{i}\b^i_\n)
\label{appo3}
\ee
and choosing $\nu_{\omega,\alpha,\alpha',i}$ so that $\nu_{-\infty,\omega,\alpha,\alpha',i}=0$ we get
\be
\nu_{h,\omega,\alpha,\alpha',i}=-\g^{-h}\sum_{i=-\infty}^{h} \g^{i}\b^i_\n
\label{appo4}
\ee
and by using a fixed point argument we can show that there is a sequence such that
$|\nu_{h,\omega,\alpha,\alpha',i}|\le C \l\g^{h\over 2}$.

The application of the above bounds to the 2-point function, in order to derive
\pref{43}, is straightforward. The 2-point function can be written as 
\be
\hat S_{j,j}(\mathbf k+K^j_\o )) =\sum_{h=-\infty}^0 
\hat g^{(h)}(\mathbf k+\mathbf K^j_\o )+r^h(\mathbf k) 
\ee
where $r^h(\mathbf k) $ includes the contribution of term with at least a vertex. We can replace in $g^{(h)}((\mathbf k+\mathbf K^j_\o ))$ the 
$v_{i,\omega,h},w_{i,\omega,h},Z_{i,\omega,h}$  with
$v_{i,\omega,-\infty},w_{i,\omega,-\infty},Z_{i,\omega,-\infty}$ obtaining the dominant term in 
\pref{43}; the subdominant term is obtained both from the 
term containing the difference between 
$v_{i,\omega,h}-v_{i,\omega,-\infty}$, $z_{i,\omega,h}-w_{i,\omega,-\infty}$,
$Z_{i,\omega,h}-Z_{i,\omega,-\infty}$, which have an extra factor $O(\l \g^{h/2})$, or the 
terms with at least a vertex 
which have at least
a $\n$ or a non resonant trivial vertex, with an extra $O(\g^{h/2})$
from the bounds after \pref{weightzz}.

\section{Conclusions }

We have considered 
the role of large-momentum-transfer Umklapp terms. These terms, which nearly connect the Fermi points of different layers, are typically neglected in effective continuum descriptions but could destroy the Dirac cones.
We rigorously find that, for finite interlayer coupling, the semimetallic phase is stable 
provided the angles belong to a fractal set of large relative measure (decreasing with the hopping)
characterized by the validity of a Diophantine condition. 
Our result therefore ensures that for weak hopping for most (in the sense specified above) angles the semimetallic behaviour persists
even in presence of the large momentum Umklapp terms which almost connect the Dirac points.
The result provides a partial justification
to the effective continuum description of TBG in which terms are neglected. 

It would be of course a very interesting problem
to consider hopping outside the regime we are considering to see if there can be some transition to a different phase due to such terms,
as it happens for fermions in a 1D quasi-periodic Aubry potential; or to detect in the weak coupling regime some sign of fractality 
following from the above number theoretical considerations.
In addition,
the present analysis paves the way to a more accurate evaluation of the
velocities as functions of the angles, taking into account lattice or higher-order effects, and the effect
of many-body interactions,
whose interplay with the emerging quasi-periodicity
could lead to interesting phases.

\begin{acknowledgements}
We thank J. Pixley for many interesting discussions.
V.M. acknowledges support from the MUR, PRIN 2022 project MaIQuFi cod. 20223J85K3.
I.J. gratefully acknowledges support through NSF Grant DMS-2349077, and the Simons Foundation, Grant Number 825876.
\end{acknowledgements}

\vfill
\pagebreak
\widetext

\appendix

\section{Fourier transform of the interlayer hopping}\label{app:fourierV}
We write $V$ in Fourier space: we get
\begin{eqnarray}
&&V=\frac \lambda{|\hat{\mathcal L}_1|^2}\sum_{x_1\in \hat{\mathcal L}_1} \sum_{x'_2\in\Lambda_2}\sum_{\alpha\in\{a,b\}}
    \int_{\mathbb R^2} \frac{dq}{4\pi^2}
    \int_{\hat{\mathcal L}_1}dk_1
    \int_{\hat{\mathcal L}_2}dk_2'\nonumber\\
&&
    e^{i(k_1x_1-k_2'x_2'+q(x_1-x_2'))}
    e^{iq(d_\alpha-Rd_{\alpha})}
    e^{i \xi (k_2'-k_1)}
    \hat\varsigma(q)\hat c^\dagger_{1,k_1,\alpha}\hat c_{2,k_2',\alpha}+\nonumber\\
&&e^{-i(k_1x_1-k_2'x_2'-q(x_1-x_2'))}
    e^{iq(d_\alpha-Rd_{\alpha})}
    e^{-i\xi(k_2'-k_1)}
    \hat\varsigma(q)\hat c^\dagger_{2,k_2',\alpha}\hat c_{1,k_1,\alpha}\label{V211}
\end{eqnarray}
and using the Poisson summation formula
\begin{equation}
  \sum_{x_1\in\mathcal L_1} e^{i (k_1+q)x_1}=|\hat{\mathcal L}_1|\sum_{l\in\mathbb Z^2} \d(k_1+q+l b)
\end{equation}
where we use the shorthand $lb\equiv l_1b_1+l_2b_2$, and
\begin{equation}
  \sum_{x_2'\in \mathcal L_2} e^{-i (k_2+q) x'_2}=|\hat{\mathcal L}_1|\sum_{m\in\mathbb Z^2} \d(k_2+q+m b')
\end{equation}
Noting that
$\hat c_{2,k_1+lb-mb',\alpha}
  \equiv
  \hat c_{2,k_1+lb,\alpha}$, $
  \hat c_{1,k_2'+mb'-lb,\alpha}
  \equiv
  \hat c_{1,k_2'+mb',\alpha}$ we finally obtain \pref{11}.
\bigskip

Rewriting (\ref{V211}) in terms of Grassmann variables, with the added imaginary time component, reads
\be\begin{array}{>\displaystyle l}
V=\frac{\beta\lambda}{|\hat\Lambda_1|^2}\sum_{x_1\in\mathcal L_1} 
\sum_{x'_2\in\mathcal L_2}\sum_{\alpha\in\{a,b\}}
\int_{\mathbb R^2} \frac{dq}{4\pi^2}
\int_{\hat\Lambda_1}d\kk_1
\int_{\hat\Lambda_2}d\kk_2'\ 
\delta(k_{1,0}-k_{2,0}')
\cdot\\[0.5cm]\cdot
\left(
e^{i(k_1x_1-k_2x_2'+q(x_1-x_2'))}
e^{iq(d_\alpha-Rd_{\alpha})}
e^{i\xi(k_2'-k_1)}
\hat\varsigma(q)\hat\psi^+_{1,\kk_1,\alpha}\hat\psi^-_{2,\kk_2',\alpha}
\label{V2112}+\right.\\[0.5cm]\indent\left.+
e^{-i(k_1x_1-k_2x_2'-q(x_1-x_2'))}
e^{iq(d_\alpha-Rd_{\alpha})}
e^{-i\xi(k_2'-k_1)}
\hat\varsigma(q)\hat\psi^+_{2,\kk_2',\alpha}\hat\psi^-_{1,\kk_1,\alpha}
\right)
\end{array}\ee
where we use the notation $\mathbf k_1=(k_{1,0},k_1)$ and $\mathbf k_2'=(k_{2,0},k_2)$.
Again, using the Poisson formula, we find \pref{112}.

\section{Proof of Lemma \ref{lemm:dioph}}\label{sec:dioph}

To prove lemma \ref{lemm:dioph}, we will first prove a general result on a Diophantine condition for a generic function from $[0,2\pi)$ to $\mathbb R^2$.
We will then apply this result to $|K^i_\omega,K^j_{\omega'}+lb+mb'|$, viewed as a function of $\theta$, for the various values of $i,j,\omega,\omega'$.

\subsection{Diophantine condition from $\mathbb R$ to $\mathbb R^2$}
Let us consider an interval $[\theta_0,\theta_1]\subset[0,2\pi]$, and define, given constants $C_1>0,\tau>4$ that are fixed once and for all, two twice-differentiable functions $x:[0,2\pi)\to \mathbb R,f:[0,2\pi)\to \mathbb R^2$, and a subset $\Omega(x,f)\subset[\theta_0,\theta_1]$,
\begin{equation}\label{diophantine}
\mathcal D(x,f):=
\{\theta\in \Omega(x,f):\ \forall k\in\mathbb Z^2\setminus\{0\},
\ \forall l\in\mathbb Z,\ |x(\theta)+l
+k \cdot f(\theta)|\geqslant C_1|k|^{-\tau}\}
\end{equation}
We will show that, provided $\Omega$ is chosen appropriately, under certain conditions on $f$ and $x$, $\mathcal D$ has a large measure.
The novelty of this result is that $f$ takes values in $\mathbb R^2$, but is a function of a single variable; if $f$ were a function from $\mathbb R^n$ to $\mathbb R^n$, then the fact that $\mathcal D$ has a large measure would follow from standard arguments.
Our result is stated for $\mathbb R^2$, but it could easily be adapted to any other dimension, provided $f$ takes a single real-valued argument.
\bigskip

In order to make our argument work, we will assume that $f'(\theta)$ (the derivative of $f$) remains inside a cone, that is, we assume that $\exists\xi\in \mathbb R^2$ with $|\xi|=1$ and $\alpha\in[0,\frac\pi4)$ such that, $\forall \theta\in [\theta_0,\theta_1]$,
\begin{equation}
  f'(\theta)\in \mathcal C_\xi(\alpha)
  :=\{y\in \mathbb R^2,\ |y\cdot\xi|>|y|\cos(\alpha)\}
  .
  \label{incone}
\end{equation}
We take the set $\Omega(x,f)$ in (\ref{diophantine}) to be
\begin{equation}
\Omega(x,f):=
\{\theta\in [\theta_0,\theta_1]:\ \forall k\in \zeta,
\ |x'(\theta)
+k \cdot f'(\theta)|\ge C_3|f'(\theta)||k|^{-\epsilon}\}
\label{Omega}
\end{equation}
where $C_3>0$ is a constant, $\epsilon\in(1,\tau-3)$, and
\begin{equation}
  \zeta:=\mathbb Z^2\setminus(\{0\}\cup\mathcal C_\xi(\beta))
  \label{zeta}
\end{equation}
where $\beta\in(\alpha,\frac\pi2-\alpha)$ (the reason why we choose $\Omega$ in this way will become apparent in the proof of Lemma \ref{lemma:diophantine} below).

\begin{lemma}\label{lemma:diophantine}
  If the following estimates hold:
  \begin{equation}
    \min_{\theta\in [\theta_0,\theta_1]}|f'(\theta)|>0
    ,\ 
    \min_{\theta\in [\theta_0,\theta_1]}|{\textstyle\frac{\partial}{\partial\theta} (\frac{f'(\theta)}{|f'(\theta)|}})|>0
    \label{bound_df}
  \end{equation}
  $\forall \theta\in [\theta_0,\theta_1]$,
  \begin{equation}
    \min_{0<|k|<R_1}|x'(\theta)+k\cdot f'(\theta)|-C_3|f'(\theta)||k|^{-\epsilon}>0
    \label{small_dx}
  \end{equation}
  with
  \begin{equation}
    R_1:= \frac{C_3+\frac{|x'(\theta)|}{|f'(\theta)|}}{\cos(\alpha+\beta)}
    \label{R1}
  \end{equation}
  and, for some $\eta>0$,
  \begin{equation}
    \min_{0<|k|<R_2}
    |{\textstyle\frac{\partial}{\partial \theta}(\frac{x'(\theta)}{|f'(\theta)|})}+k\cdot {\textstyle\frac \partial{\partial \theta}(\frac{f'(\theta)}{|f'(\theta)|})}|
    -\eta|k||{\textstyle\frac \partial{\partial \theta}(\frac{f'(\theta)}{|f'(\theta)|})}|\cos(\alpha+{\textstyle\frac\pi2}-\beta)
    >0
    \label{small_ddx}
  \end{equation}
  with
  \begin{equation}
    R_2:=
    \eta+\frac{|{\textstyle \frac \partial{\partial \theta}(\frac{x'(\theta)}{|f'(\theta)|})}|}{|{\textstyle\frac \partial{\partial \theta}(\frac{f'(\theta)}{|f'(\theta)|})}|\cos(\alpha+\frac\pi2-\beta)}
    \label{R2}
  \end{equation}
  then the measure of the complement of $\mathcal D$ is bounded by
  \begin{equation}
    |[\theta_0,\theta_1]\setminus\mathcal D(x,f)|
    \le
    O(C_3)+O({\textstyle \frac{C_1}{C_3}})
  \end{equation}
  where the constants in $O(\cdot)$ depend only on $\theta_0$, $\theta_1$, $x$, $f$, $\alpha$, $\beta$, $\epsilon$, $\tau$, $\eta$.
  In particular, if we choose $C_3\ll \theta_1-\theta_0$ and $C_1\ll (\theta_1-\theta_0)^2$, then $\mathcal D(x,f)$ fills most of $[\theta_0,\theta_1]$.
  If we further choose $C_3\sim \sqrt{C_1}\ll(\theta_1-\theta_0)$, then the measure of the complement of $\mathcal D(x,f)$ is $O(\sqrt{C_1})\ll \theta_1-\theta_0$.
\end{lemma}

\begin{remark}
  The conditions (\ref{small_dx}) and (\ref{small_ddx}) concern a finite number of values of $k$.
  In the applications of this lemma below, we can make both of these conditions trivial by ensuring that $R_1,R_2<1$, which reduces this finite number of values for $k$ to $0$.
\end{remark}

{\it Proof}
Let 
  $|\mathcal D_\Omega^c(x,f)|$ denote the Lebesgue measure of the complement $\Omega(x,f)\setminus\mathcal D(x,f)$.
  Let
  \begin{equation}
    g_{l,k}(\th)=x(\theta)+l+k \cdot f(\theta)
  \end{equation}
  in terms of which
  \begin{equation}
    |\mathcal D_\Omega^c(x,f)|\le\sum_{k,l}^* \int_{-C_1|k|^{-\tau}}^{C_1|k|^{-\tau}} \frac1{g_{l,k}'}dg_{l,k}
  \end{equation}
  where $\sum^*_{k,l}$ has the constraint that $\exists \theta\in[\theta_0,\theta_1]$ such that $g_{l,k}(\theta)\in[-C_1|k|^{-\tau},C_1|k|^{-\tau}]$. Therefore 
  \be
  |\mathcal D_\Omega^c(x,f)|\le
  \sum^*_{l,k} 2C_1 {|k|^{-\tau}\over \min_{\theta\in \Omega(x,f)} |x'(\th)+k\cdot f'(\theta)|}
  .
  \ee
    In addition, the number of values of $l$ such that $g_{l,k}(\theta)\in[-C_1|k|^{-\tau},C_1|k|^{-\tau}]$ is bounded by $C_2|k|$ for some constant $C_2$ (which depends only on $\theta_0,\theta_1,x,f$), and so
  \be
  |\mathcal D_\Omega^c(x,f)|\le
  2\sum_{k\in \mathbb Z^2\setminus\{0\}} C_2 C_1 {|k|^{1-\tau}\over \min_{\theta\in \Omega(x,f)} |x'(\th)+k\cdot f'(\theta)|} 
  .
  \label{bound_Dctmp}\ee
  In order for this bound to be useful, we must obtain a good lower bound on $|x'+k\cdot f'|$.

  To do so, $\Omega$ must be chosen appropriately: we wish for $k\cdot f'$ to stay as far away from $-x'$ as possible.
  Now, it cannot avoid it entirely, as $k\cdot f'$ will cover all possible values as $k$ varies in $\mathbb Z^2\setminus\{0\}$.
  By choosing $\Omega$ as in (\ref{Omega}), we ensure that $k\cdot f'$ may only approach $-x'$ for large values of $k$.
  In doing so, we can estimate $\mathcal D_\Omega^c$: we split the sum over $\mathbb Z^2\setminus\{0\}$ into a sum over $\zeta$ and a sum over its complement $\zeta^c\equiv\mathbb Z^2\cap\mathcal C_\xi(\beta)$, and compute a bound for each case.

  If $k\in \zeta$, then, by (\ref{Omega}), for $\theta\in \Omega(x,f)$,
  \begin{equation}
    |x'(\theta)+k\cdot f'(\theta)|\ge C_3|f'(\theta)||k|^{-\epsilon}
    .
    \label{boundin}
  \end{equation}
  If, on the other hand, $k\in \zeta^c\equiv\mathbb Z^2\cap\mathcal C_\xi(\beta)$,
  \begin{equation}
    |k\cdot f'(\theta)|\ge|k||f'(\theta)|\cos(\alpha+\beta)
  \end{equation}
  so
  \begin{equation}
    |x'(\theta)+k\cdot f'(\theta)|\ge|k||f'(\theta)|\cos(\alpha+\beta)-|x'(\theta)|
    .
  \end{equation}
  We distinguish two cases once more: either
  \begin{equation}
    |k|\ge \frac{C_3+\frac{|x'(\theta)|}{|f'(\theta)|}}{\cos(\alpha+\beta)}\equiv R_1
  \end{equation}
  (see (\ref{R1})) in which case (\ref{boundin}) holds true for these $k$'s as well, or $|k|<R_1$, in which case (\ref{boundin}) holds by virtue of (\ref{small_dx}).
  All in all, whatever the value of $k$, (\ref{boundin}) holds for all $k\in \mathbb Z^2\setminus\{0\}$.

  Therefore, (\ref{bound_Dctmp}) becomes
  \be
  |\mathcal D_\Omega^c(x,f)|\le
  \frac{2C_1C_2}{C_3\min_{\theta\in [\theta_0,\theta_1]} |f'(\theta)|}\sum_{k\in \mathbb Z^2\setminus\{0\}} |k|^{1-\tau+\epsilon} 
  .
  \label{bound_D}\ee
  By (\ref{bound_df}), this sum is bounded since $\epsilon <\tau-3$.

  We are left with estimating the measure of $\Omega^c(x,f):=[\theta_0,\theta_1]\setminus \Omega(x,f)$.
  Proceeding in a similar way as for $|\mathcal D_\Omega^c(x,f)|$, we find that
  \begin{equation}
    |\Omega^c(x,f)|\le 2C_3\sum_{k\in \zeta}\frac{|k|^{-\epsilon}}{\min_{\theta\in[\theta_0,\theta_1]}|\frac{\partial}{\partial \theta}( \frac{x'(\theta)}{|f'(\theta)|})+k\cdot \frac\partial{\partial \theta}(\frac{f'(\theta)}{|f'(\theta)|})|}
    .
  \end{equation}
  Here we see why it was necessary to introduce the set $\zeta$ in (\ref{Omega}): if $\zeta$ were taken to be $\mathbb Z^2\setminus\{0\}$, then we would run into exactly the same problem as before: the denominator in this bound would not be bounded away from 0.
  However, the problem that gave rise to the necessity of introducing $\Omega$ in the first place actually only occurred for the $k$'s that are close to being orthogonal to $f'(\theta)$.
  So we can restrict $\zeta$ to only include the $k$'s that are close to being orthogonal to $f'(\theta)$, which is why we define $\zeta$ as in (\ref{zeta}).
  Because $\frac\partial{\partial \theta}( \frac{f'}{|f'|})$ is orthogonal to $f'$,
  \begin{equation}
    {\textstyle\frac\partial{\partial \theta}(\frac{f'}{|f'|})}\in \mathcal C_{\xi^\perp}(\alpha)
  \end{equation}
  and so, for $k\in \zeta$, because the maximal angle between $k$ and ${\textstyle\frac\partial{\partial \theta}(\frac{f'}{|f'|})}$ is $\alpha+\frac\pi2-\beta$,
  \begin{equation}
    |k\cdot {\textstyle\frac \partial{\partial \theta}(\frac{f'(\theta)}{|f'(\theta)|})}|>
    |k||{\textstyle\frac \partial{\partial \theta}(\frac{f'(\theta)}{|f'(\theta)|})}|\cos(\alpha+{\textstyle\frac\pi2}-\beta)
    .
  \end{equation}
  Thus,
  \begin{equation}
    |{\textstyle\frac \partial{\partial \theta}(\frac{x'(\theta)}{|f'(\theta)|})}+k\cdot {\textstyle\frac \partial{\partial \theta}(\frac{f'(\theta)}{|f'(\theta)|})}|
    >
    |k||{\textstyle\frac \partial{\partial \theta}(\frac{f'(\theta)}{|f'(\theta)|})}|\cos(\alpha+{\textstyle\frac\pi2}-\beta)
    -|{\textstyle\frac \partial{\partial \theta}(\frac{x'(\theta)}{|f'(\theta)|})}|
    .
  \end{equation}
  Therefore, if
  \begin{equation}
    |k|\ge
    \eta+\frac{|{\textstyle \frac \partial{\partial \theta}(\frac{x'(\theta)}{|f'(\theta)|})}|}{|{\textstyle\frac \partial{\partial \theta}(\frac{f'(\theta)}{|f'(\theta)|})}|\cos(\alpha+{\textstyle\frac\pi2}-\beta)}
    \equiv R_2
  \end{equation}
  then
  \begin{equation}
    |{\textstyle\frac \partial{\partial \theta}(\frac{x'(\theta)}{|f'(\theta)|})}+k\cdot {\textstyle\frac \partial{\partial \theta}(\frac{f'(\theta)}{|f'(\theta)|})}|
    >\eta|k||{\textstyle\frac \partial{\partial \theta}(\frac{f'(\theta)}{|f'(\theta)|})}|\cos(\alpha+{\textstyle\frac\pi2}-\beta)
    .
    \label{tmpineq}
  \end{equation}
  If, on the other hand, $|k|<R_2$, then (\ref{tmpineq}) holds by virtue of (\ref{small_ddx}).
  Thus, (\ref{tmpineq}) holds for all $k\in \zeta$.
  Therefore,
  \begin{equation}
    |\Omega^c(x,f)|
    \le
    \frac{2C_3}{\eta{\displaystyle\min_{\theta\in[\theta_0,\theta_1]}}|\frac\partial{\partial \theta}(\frac{f'(\theta)}{|f'(\theta)|})|\cos(\alpha+\frac\pi2-\beta)}
    \sum_{k\in \mathbb Z^2\setminus\{0\}}|k|^{-1-\epsilon}
    .
    \label{bound_Omega}
  \end{equation}
  By (\ref{bound_df}), this sum is bounded since $\epsilon>1$.
  We conclude the proof by combining (\ref{bound_D}) with (\ref{bound_Omega}).
\qed

\subsection{Applying the Diophantine condition to $|K^i_\omega-K^j_{\omega'}+lb+mb'|$}\label{app:dioph}
We now apply lemma \ref{lemma:diophantine} repeatedly to prove lemma \ref{lemm:dioph}.
Throughout this section, we set $\beta=\frac\pi4$ and take $\alpha$ very small.
Our proof is computer assisted.

Let us first consider the case $i=j$, $\omega=\omega'$, and $y=m$, that is, we wish to find a condition on $\theta$ such that
\begin{equation}
  |K^i_\omega-K^j_{\omega'}+lb+mb'|
  \equiv |lb+mb'|
  \equiv |M|
  \ge\frac{C_0}{|m|^\tau}
  \label{cond11}
\end{equation}
(recall (\ref{cond}) and (\ref{M})).
We recall (\ref{Mineq}):
\begin{equation}
  |M|\ge
  \frac{2\pi}{\sqrt3}(l_\omega+m\cdot f_\omega)
\end{equation}
with $l_+\equiv l_1$ and $l_-\equiv l_2$, and
\begin{equation}
  f_\omega(\theta):=({\textstyle \frac{\varphi_1-\omega\varphi_3}2},\ {\textstyle \frac{\varphi_2+\omega\varphi_4}2})
  .
\end{equation}
Now, recalling the definition (\ref{diophantine}), we have that if $\theta\in \mathcal D(0,f_\omega)$, then the inequality (\ref{cond}) with $i=i'$, $\omega=\omega'$, and $y=m$ holds with $C_0:=\frac{2\pi}{\sqrt3}C_1$.
We therefore just need to use Lemma \ref{lemma:diophantine} to ensure that the measure of this set is large.
Taking $\theta_0$ and $\theta_1$ to be sufficiently close to each other (so that the value of $\alpha$ is arbitrarily small and the cone $\mathcal C_\xi(\alpha)$ is arbitrarily narrow), it suffices to verify the conditions at $\alpha=0$, and conclude by continuity.

\begin{figure}
  \includegraphics[width=6cm]{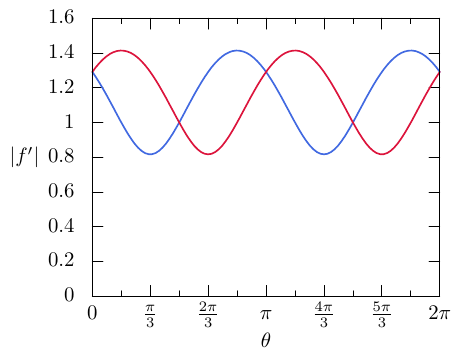}
  \includegraphics[width=6cm]{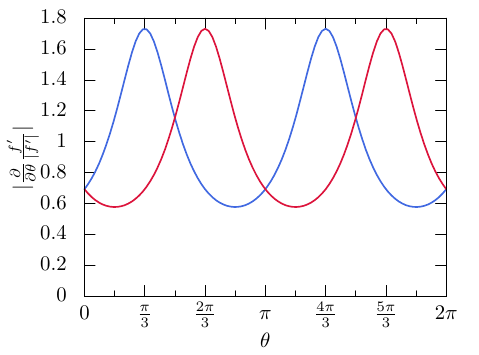}

  \caption{$|f'_\omega(\theta)|$ and $|\frac \partial{\partial \theta}\frac{f'_\omega(\theta)}{|f'_\omega(\theta)|}|$ (blue: $\omega=+$, red: $\omega=-$, color online).
  These show (\ref{bound_df}) holds: neither of these quantities reaches 0.}
  \label{fig:df}
\end{figure}

We first verify (\ref{bound_df}).
In Figure \ref{fig:df}, we see that neither $f'$ nor $\frac \partial{\partial \theta}\frac{f'_\omega(\theta)}{|f'_\omega(\theta)|}$ vanish for any value of $\theta\in[0,2\pi)$, so (\ref{bound_df}) holds.

Since $x=0$, the other assumptions trivially hold: we choose $C_3<1/\sqrt2$ and $\eta<1$ such that $R_1,R_2<1$, in which case the minima in (\ref{small_dx}) and (\ref{small_ddx}) are taken over empty sets, so (\ref{small_dx}) and (\ref{small_ddx}) hold trivially.
Thus, by Lemma \ref{lemma:diophantine}, choosing $C_1\ll(\theta_1-\theta_0)^2$, the set $\mathcal D(0,f_\omega)$ has a large measure.

We now repeat the argument for the other values of $i,j$, $y$, and $\omega,\omega'$.
First, note that $K_+-K_-=\frac13(b_1-b_2)$ so the condition (\ref{cond}) holds for $\o\neq \o'$ whenever it holds for $\omega=\omega'$.
Next, note that
\begin{equation}
  |K^i_\omega-K^j_\omega+lb+mb'|
  =
  |R^T(K^i_\omega-K^j_\omega+lb+mb')|
\end{equation}
which corresponds to exchanging $m$ and $l$, and flipping the sign of $\theta$.
The arguments made for $\theta$ may be adapted in a straightforward way to the case $-\theta$ so  our derivation for $y=m$ also applies to $y=l$.

We are thus left with the case $i\neq j$, $\omega=\omega'$, and $y=m$.
Without loss of generality, we choose $i=1$, $j=2$, and we wish to bound
\begin{equation}
  |K^1_\omega-K^2_{\omega}+lb+mb'|
  \ge\frac{C_0}{|m|^\tau}
  \label{cond12}
\end{equation}
Proceeding as we did above, we bound
\begin{equation}
  |K^1_\omega-K^2_{\omega}+lb+mb'|
  \ge
  \frac{2\pi}{\sqrt3}\left(
    x_\omega(\theta)
    +l_1+m\cdot f_\omega(\theta)
  \right)
\end{equation}
with
\begin{equation}
  x_\omega(\theta):=\frac{1}3(1-c_\theta)+\omega\frac{1}{\sqrt3}s_\theta
  .
\end{equation}
Therefore, if $\theta\in \mathcal D(x_\omega,f_\omega)$, then (\ref{cond12}) holds with $C_0=\frac{2\pi}{\sqrt3}C_1$.
To show that this set has a large measure, we check the assumptions of Lemma \ref{lemma:diophantine}, as we did above.
Again, we check the assumptions at $\alpha=0$, and argue by continuity.

\begin{figure}
  \includegraphics[width=6cm]{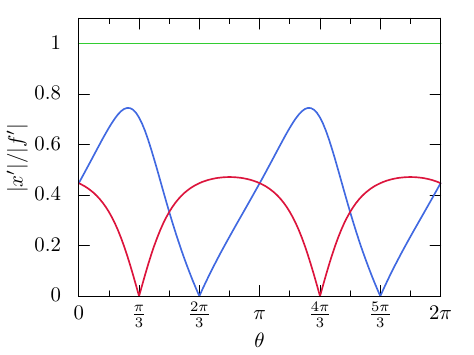}
  \includegraphics[width=6cm]{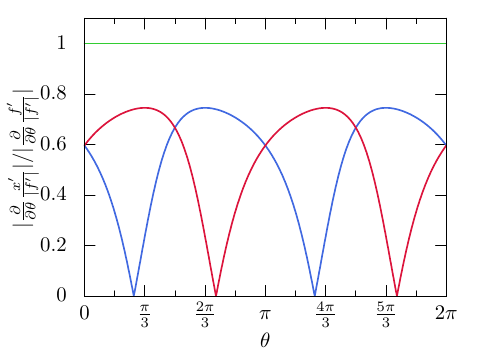}

  \caption{$\frac{|x'_\omega(\theta)|}{|f'_\omega(\theta)|}$ and $|\frac \partial{\partial \theta}\frac{x'_\omega(\theta)}{|f'_\omega(\theta)|}|/|\frac \partial{\partial \theta}\frac{f'_\omega(\theta)}{|f'_\omega(\theta)|}|$ (blue: $\omega=+$, red: $\omega=-$, color online).
  The green lines represent the value 1.
  The plot on the left is used to verify (\ref{small_dx}): as we see, $\frac{|x_\omega'(\theta)|}{|f_\omega'(\theta)|}<1$, so, to verify (\ref{small_dx}), it suffices to check the cases for which $|k|=1$.
  The plot on the right is used to verify (\ref{small_ddx}): as we see, $|\frac \partial{\partial \theta}\frac{x'_\omega(\theta)}{|f'_\omega(\theta)|}|/|\frac \partial{\partial \theta}\frac{f'_\omega(\theta)}{|f'_\omega(\theta)|}|<1$, so to verify (\ref{small_ddx}), it suffices to check the cases for which $|k|=1$.}
  \label{fig:ddx}
\end{figure}

\begin{figure}
  \includegraphics[width=6cm]{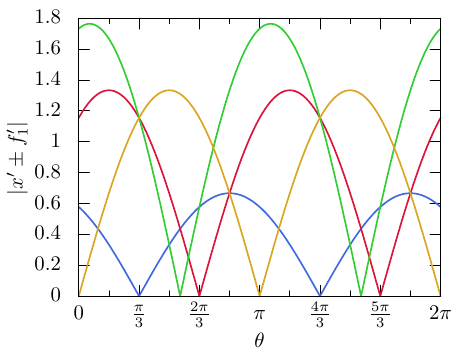}
  \includegraphics[width=6cm]{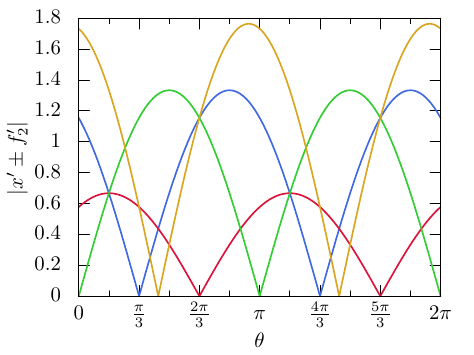}

  \caption{Left: $|x'\pm f_1'|$, right: $|x'\pm f_2'|$, in each graph, blue and green are $\omega=+$, red and yellow are $\omega=-$, blue and red are $\pm=+$, green and yellow are $\pm=-$.
  These plots are used to verify (\ref{small_dx}) for $|k|=1$.
  As we see, $x'+k \partial f$ only reaches 0 at a finite number of points, enumerated in (\ref{B1}).}
  \label{fig:resonance1}
\end{figure}

We check (\ref{small_dx}).
We choose $C_3$ in such a way that $R_1<\sqrt2$.
In Figure \ref{fig:ddx} (left), we see that $\frac{|x_\omega'(\theta)|}{|f_\omega'(\theta)|}<1$ for all $\theta\in[0,2\pi)$, so we choose $C_3$ such that  $C_3<1-\frac{|x'_\omega(\theta)|}{|f'_\omega(\theta)|}$ (recall that, since $\beta=\frac\pi4$, $\sqrt2\cos(\beta)=1$).
Thus, to verify (\ref{small_dx}), it suffices to check the cases for which $|k|=1$, that is, $k=(\pm 1,0)$ and $k=(0,\pm1)$.
And, again choosing $C_3$ to be small enough, it suffices to check that $x'_\omega(\theta)+k\cdot f'_\omega(\theta)\neq 0$.
Using a symbolic calculator, we find that this holds as long as
\begin{equation}
  \theta\not\in
  \mathcal B_1:=\left\{
    0,
    \frac\pi3,
    \arctan(3\sqrt3),
    \pi-\arctan(3\sqrt3),
    \frac{2\pi}3,
    \pi,
    \frac{4\pi}3,
    \pi+\arctan(3\sqrt3),
    2\pi-\arctan(3\sqrt3),
    \frac{5\pi}3
  \right\}
  .
  \label{B1}
\end{equation}
(Figure \ref{fig:resonance1} shows the plot of $x'_\omega(\theta)+k\cdot f'_\omega(\theta)$.)

\begin{figure}
  \includegraphics[width=6cm]{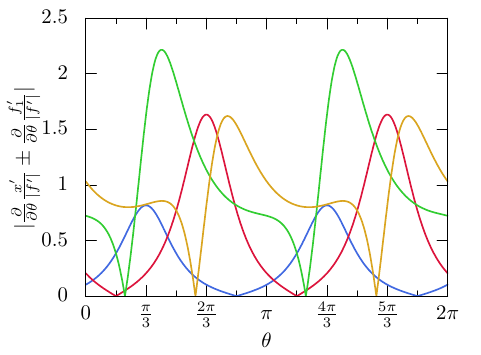}
  \includegraphics[width=6cm]{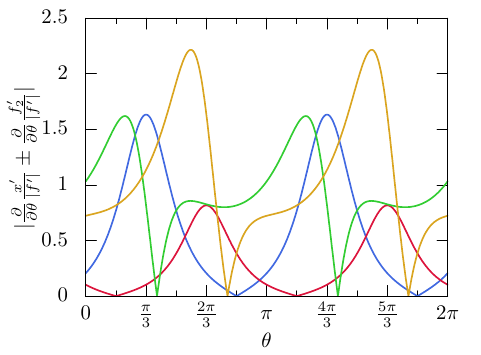}

  \caption{Left: $|\frac\partial{\partial \theta}\frac{x'_\omega}{|f'_\omega|}\pm\frac\partial{\partial \theta}\frac{f'_{\omega,1}}{|f'_\omega|}|$, right: $|\frac\partial{\partial \theta}\frac{x'_\omega}{|f'_\omega|}\pm\frac\partial{\partial \theta}\frac{f'_{\omega,2}}{|f'_\omega|}|$, in each graph, blue and green are $\omega=+$, red and yellow are $\omega=-$, blue and red are $\pm=+$, green and yellow are $\pm=-$.
  These plots are used to verify (\ref{small_ddx}) for $|k|=1$.
  As we see, $\frac\partial{\partial \theta}\frac{x'_\omega}{|f'_\omega|}+k\cdot\frac\partial{\partial \theta}\frac{f'_\omega}{|f'_\omega|}$ only reaches 0 at a finite number of points, enumerated in (\ref{B2}).}
  \label{fig:resonance3}
\end{figure}

Finally, we check (\ref{small_ddx}).
We choose $\eta$ in such a way that $R_2<\sqrt2$.
In Figure \ref{fig:ddx} (right), we see that $\left|\frac\partial{\partial \theta}\frac{x'_\omega}{|f'_\omega|}\right|<\left|\frac\partial{\partial \theta}\frac{f'_\omega}{|f'_\omega|}\right|$ for all $\theta\in[0,2\pi)$, so we choose $\eta$ such that $\eta<\sqrt2-\sqrt2\left|\frac\partial{\partial \theta}\frac{x'_\omega}{|f'_\omega|}\right|/\left|\frac\partial{\partial \theta}\frac{f'_\omega}{|f'_\omega|}\right|$ (recall that, since $\beta=\frac\pi4$, $\cos(\frac\pi2-\beta)=\frac1{\sqrt2}$).
Thus, to verify (\ref{small_ddx}), it suffices to check the cases for which $|k|=1$, that is, $k=(\pm 1,0)$ and $k=(0,\pm1)$.
And, again choosing $\eta$ to be small enough, it suffices to check that $\frac\partial{\partial \theta}\frac{x'_\omega}{|f'_\omega|}+k\cdot\frac\partial{\partial \theta}\frac{f'_\omega}{|f'_\omega|}\neq0$.
Using a symbolic calculator, we find that this holds as long as
\begin{equation}
  \begin{array}{>\displaystyle l}
  \theta\not\in
  \mathcal B_2:=\left\{
    \frac{\pi}6,
    \arccos({\textstyle\frac{5\sqrt3}{2\sqrt{31}}}),
    \arccos({\textstyle\frac{\sqrt3}{2\sqrt{7}}}),
    \pi-\arccos({\textstyle\frac{\sqrt3}{2\sqrt{7}}}),
    \pi-\arccos({\textstyle\frac{5\sqrt3}{2\sqrt{31}}}),
    \frac{5\pi}6,
    \right.\\\hskip3cm\left.
    \frac{7\pi}6,
    \pi+\arccos({\textstyle\frac{5\sqrt3}{2\sqrt{31}}}),
    \pi+\arccos({\textstyle\frac{\sqrt3}{2\sqrt{7}}}),
    2\pi-\arccos({\textstyle\frac{\sqrt3}{2\sqrt{7}}}),
    2\pi-\arccos({\textstyle\frac{5\sqrt3}{2\sqrt{31}}}),
    \frac{11\pi}6
  \right\}
  .
  \end{array}
  \label{B2}
\end{equation}
(Figure \ref{fig:resonance3} shows the plot of $\frac\partial{\partial \theta}\frac{x'_\omega}{|f'_\omega|}+k\cdot\frac\partial{\partial \theta}\frac{f'_\omega}{|f'_\omega|}$.)

All in all, we have found that if we restrict the values of $\theta$ to a small interval $[\theta_0,\theta_1]$ that steers clear of any of the angles in $\mathcal B_1\cup \mathcal B_2$ (see (\ref{B1}) and (\ref{B2})), and if we further restrict $\theta$ to an intersection of Diophantine sets:
\begin{equation}
  \theta\in[\theta_0,\theta_1]\cap
  \bigcap_{\omega=\pm}\bigcap_{\sigma=\pm}\mathcal D(0,f_\omega(\sigma \theta))
  \cap
  \bigcap_{\omega=\pm}\bigcap_{\sigma=\pm}\mathcal D(x_\omega(\sigma \theta),f_\omega(\sigma \theta))
\end{equation}
then (\ref{cond}) is satisfied for any value of $i,i'$, $\omega,\omega'$, and $y$ with a constant $C_0=O((\theta_1-\theta_0)^2)$.
Because each set has an arbitrarily large measure (relative to $[\theta_0,\theta_1]$), their intersection also does.

\section{Naive perturbation theory} \label{app:explicit_feynman}

The {\it 2-point function} function is computed using perturbation theory: formally,
\begin{equation}
\begin{array}{>\displaystyle l}
  (S_{1,1}(\mathbf k))_{\alpha',\alpha}=
  \sum_{N=0}^\infty
  \sum_{\alpha_0,\cdots,\alpha_{2N+1}}
  (g_1(\mathbf k))_{\alpha,\alpha_0}
  \left(\prod_{n=0}^N
  \left(
    \sum_{l_{2n}}\tau_{l_{2n},\alpha_{2n}}^{(1)}(k+m_{2n-1}b'+l_{2n}b)
    (g_2(\mathbf k+l_{2n}b))_{\alpha_{2n},\alpha_{2n+1}}
    \cdot\right.\right.\\\hfill\cdot\left.\left.
    \sum_{m_{2n+1}}\tau_{m_{2n+1},\alpha_{2n+1}}^{(2)}(k+l_{2n}b+m_{2n+1}b')
    ((g_1(\mathbf k+m_{2n+1}b'))_{\alpha_{2n+1},\alpha_{2n+2}})^{\mathds 1_{n<N}}
  \right)\right)
  (g_1(\mathbf k))_{\alpha_{2N+1},\alpha'}
  \label{linear1}
\end{array}\end{equation}
where $m_{-1}\equiv l_{-1}\equiv0$ and $\mathds 1_{n<N}\in\{0,1\}$ is equal to 1 if and only if $n<N$,
\begin{equation}\begin{array}{>\displaystyle l}
  (S_{2,2}(\mathbf k))_{\alpha',\alpha}=
  \sum_{N=0}^\infty
  \sum_{\alpha_0,\cdots,\alpha_{2N+1}}
  (g_2(\mathbf k))_{\alpha,\alpha_0}
  \left(\prod_{n=0}^N
  \left(
    \sum_{m_{2n}}\tau_{m_{2n},\alpha_{2n}}^{(2)}(k+l_{2n-1}b+m_{2n}b')
    (g_1(\mathbf k+m_{2n}b'))_{\alpha_{2n},\alpha_{2n+1}}
    \cdot\right.\right.\\\hfill\cdot\left.\left.
    \sum_{l_{2n+1}}\tau_{l_{2n+1},\alpha_{2n+1}}^{(1)}(k+m_{2n}b'+l_{2n+1}b)
    ((g_2(\mathbf k+l_{2n+1}b))_{\alpha_{2n+1},\alpha_{2n+2}})^{\mathds 1_{n<N}}
  \right)\right)
  (g_2(\mathbf k))_{\alpha_{2N+1},\alpha'}
\end{array}\end{equation}
\begin{equation}\begin{array}{>\displaystyle l}
  (S_{2,1}(\mathbf k))_{\alpha',\alpha}=
  \sum_{N=0}^\infty
  \sum_{\alpha_0,\cdots,\alpha_{2N}}
  (g_1(\mathbf k))_{\alpha,\alpha_0}
  \left(\prod_{n=0}^N
  \left(
    \sum_{l_{2n}}\tau_{l_{2n},\alpha_{2n}}^{(1)}(k+m_{2n-1}b'+l_{2n}b)
    ((g_2(\mathbf k+l_{2n}b))_{\alpha_{2n},\alpha_{2n+1}})^{\mathds 1_{n<N}}
    \cdot\right.\right.\\\hfill\cdot\left.\left.
    \left(\sum_{m_{2n+1}}\tau_{m_{2n+1},\alpha_{2n+1}}^{(2)}(k+l_{2n}b+m_{2n+1}b')
    (g_1(\mathbf k+m_{2n+1}b'))_{\alpha_{2n+1,2n+2}}\right)^{\mathds 1_{n<N}}
  \right)\right)
  (g_2(\mathbf k))_{\alpha_{2N,\alpha'}}
\end{array}\end{equation}
\begin{equation}\begin{array}{>\displaystyle l}
  (S_{1,2}(\mathbf k))_{\alpha',\alpha}=
  \sum_{N=0}^\infty
  \sum_{\alpha_0,\cdots,\alpha_{2N}}
  (g_2(\mathbf k))_{\alpha,\alpha_0}
  \left(\prod_{n=0}^N
  \left(
    \sum_{m_{2n}}\tau_{m_{2n},\alpha_{2n}}^{(2)}(k+l_{2n-1}b+m_{2n}b')
    ((g_1(\mathbf k+m_{2n}b'))_{\alpha_{2n},\alpha_{2n+1}})^{\mathds 1_{n<N}}
    \cdot\right.\right.\\\hfill\cdot\left.\left.
    \left(\sum_{l_{2n+1}}\tau_{l_{2n+1}}^{(1)}(k+m_{2n}b'+l_{2n+1}b)
    (g_2(\mathbf k+l_{2n+1}b))_{\alpha_{2n+1},\alpha_{2n+2}}\right)^{\mathds 1_{n<N}}
  \right)\right)
  \label{linear4}
\end{array}\end{equation}

\section {Symmetry constraints on the resonant terms}\label{sec:symm}

\subsection{Symmetries of the system}\label{sec:symmetry}

Let us state the symmetries of the model, which will play an important role in our discussion.
Note that the $C_2T$ symmetry in (c) only holds if $\xi=(0,1/2)$.
\vskip.3cm
a) {\it  Complex conjugation}:
every complex constant is conjugated and
\be
\hat\psi_{1,k_0,k,\alpha}^\pm\mapsto
  e^{\mp i \xi(b_1+b_2)}
  \hat\psi_{1,-k_0, -k,\alpha}^\pm
  ,\quad
  \label{sim}
\hat\psi_{2,k_0,k,\alpha}^\pm\mapsto 
  e^{\mp i \xi(b'_1+b'_2)}
  \hat\psi_{2,-k_0,-k,\alpha}^\pm\ee
Indeed, it is straighforward to check that $H_1$ and $H_2$ (see (\ref{H1k}) and (\ref{H2k})) are left invariant by (\ref{sim}) (following from the fact that $\Omega(-k)=\Omega(k)^*$).
To check the invariance of the interlayer hopping term $V$ (see (\ref{V2112})), there is one subtelty: because $\hat{\mathcal L}_1$ and $\hat{\mathcal L}_2$ have different periodicity, we cannot simply change $k_1$ to $-k_1$ and $k_2'$ to $-k_2'$, which would not leave $\hat{\mathcal L}_i$ invariant.
Instead, we map $k_1$ to $-k_1+b_1+b_2$ and $k_2'$ to $-k_2'+b_1'+b_2'$.
It is then straightforward to check (using (\ref{V2112})) that $V$ remains invariant under (\ref{sim}), using $e^{i(1,1)bx_1}=e^{i(1,1)b'x_2'}=1$
and periodicity.
\vskip.3cm

b)
 {\it Particle-hole} \be
\hat\psi_{1,k_0, k,\alpha}^\pm\mapsto
  ie^{\pm i\xi(b_1+b_2)}
  \hat\psi_{1,k_0,-k,\alpha}^\mp
  ,\quad
  \label{sim1}
  \hat\psi_{2,k_0, k,\alpha}^\pm\mapsto 
  ie^{\pm i\xi(b'_1+b'_2)}
  \hat\psi_{2,k_0, -k,\alpha}^\mp\ee
The argument is substantially the same as for the conjugation symmetry (a).
\vskip.3cm

 c) {\it $C_2 T$ symmetry}
\be
\hat\psi_{j,k_0,k,\alpha}^\pm\mapsto
  e^{\pm i\chi_j(k)}
  \hat\psi_{j,k_0,-k,\bar\alpha}^\pm
  \label{C2T}
\ee
where if $\alpha=a$ then $\bar \alpha=b$ and if $\alpha=b$ then $\bar \alpha =a$, and
\begin{equation}
  \chi_1(k):=
  \frac{d_b}2(b_1+b_2)-kd_b-\sigma_{k,2}bd_b
  ,\quad
  \chi_2(k):=
  \frac{d_b}2(b_1'+b_2')-kRd_b-\sigma_{k,1}b'd_b
  .
\end{equation}
$H_1$ and $H_2$ are invariant under (\ref{C2T}) for the same reason as the conjugation and particle-hole symmetries.
For the interlayer hopping, it is easiest to use the expression of $V$ in (\ref{11}), which involves two integrals: one over $\hat{\mathcal L}_1$ and one over $\hat{\mathcal L}_2$.
Let us discuss the invariance of the integral over $\hat{\mathcal L}_1$, as the invariance of the other follows from a similar argument.
We change variables in the integral: $k\mapsto -k+b_1+b_2$, as well as in the sum over $l$: $(l_1,l_2)\mapsto -(l_1+1,l_2+1)$, and in the sum over $\alpha$: $\alpha\mapsto\bar \alpha$.
This changes $\tau_{l,\alpha}^{(1)}(k+lb)$ to $\tau_{-l-(1,1),\bar \alpha}^{(1)}(-k-lb)$.
Now, by (\ref{tau1}),
\begin{equation}
  \frac{\tau_{-l-(1,1),\bar \alpha}^{(1)}(-k-lb)}{\tau_{l,\alpha}^{(1)}(k+lb)}
  =
  e^{-i\xi(2l+(1,1))b}
  e^{i(k+lb)(d_{\bar \alpha}+d_\alpha-Rd_{\alpha}-Rd_{\bar \alpha})}
  e^{-i\xi (\sigma_{-k-lb,1}-\sigma_{k+lb,1})b'}
  \frac{\hat\varsigma(k+lb)}{\hat\varsigma^*(k+lb)}
  .
\end{equation}
In addition, if $k+lb-\sigma_{k+lb,1}b'\in\hat{\mathcal L}_2$, then $-k-lb+(\sigma_{k+lb,1}+(1,1))b'\in\hat{\mathcal L}_2$, so
\begin{equation}
  \sigma_{-k-lb,1}=-\sigma_{k+lb,1}-(1,1)
\end{equation}
and, since $d_a=0$ and $d_b=(1,0)$,
\begin{equation}
  d_{\bar \alpha}+d_\alpha-Rd_{\alpha}-Rd_{\bar \alpha}
  =d_b-Rd_b
\end{equation}
and, since $\varsigma(x)=\varsigma(-x)$, $\hat\varsigma\in \mathbb R$.
Therefore,
\begin{equation}
  \frac{\tau_{-l-(1,1),\bar \alpha}^{(1)}(-k-lb)}{\tau_{l,\alpha}^{(1)}(k+lb)}
  =
  e^{-i\xi(2l+(1,1))b}
  e^{i(k+lb)(d_b-Rd_b)}
  e^{i\xi (2 \sigma_{k+lb,1}+(1,1))b'}
  .
\end{equation}
Since $\xi=d_b/2$,
\begin{equation}
  \frac{\tau_{-l-(1,1),\bar \alpha}^{(1)}(-k-lb)}{\tau_{l,\alpha}^{(1)}(k+lb)}
  =
  e^{ikd_b}
  e^{-i(k+lb)Rd_b}
  e^{i d_b\sigma_{k+lb,1}b'}
  e^{i\frac{d_b}2(b_1'+b_2'-b_1-b_2)}
  .
\end{equation}
It is then straightforward to check that this extra phase gets canceled out exactly by $e^{\pm i\chi_j}$ in (\ref{C2T}) (to see this, note that if $k\in\hat{\mathcal L}_1$, then $\sigma_{k,2}=0$).

\vskip.3cm

d) {\it Inversion} 
\be \hat\psi_{j,k_0,k,\alpha}^\pm\to i (-1)^\alpha (-1)^j\hat\psi_{j,-k_0,k,\alpha}^\pm\label{inversion}\ee
\vskip.3cm 
It is straightforward to check that $H_1$, $H_2$, and the interlayer hopping (using (\ref{V2112})) are invariant under (\ref{inversion}).

\subsection {Constraints on the resonant terms}

The discrete symmetry properties seen above implies
severely constraint the form of the resonant terms. 
In the following, we use the notation ``$=^a$'' to mean ``by using symmetry (a) from Section \ref{sec:symmetry} (that is, Complex conjugation), it is equal to'', and similarly for ``$=^b$'', ``$=^c$'', ``$=^d$''.

\begin{enumerate}

\item
Using that $W_{aa}(k_0,k)=^d-W_{aa}(-k_0,k)$
we get  $W_{aa}(0,K_\omega)=\partial_1W_{aa}(0,K_\omega)=\partial_2W_{aa}(0,K_\omega)=0$.
Similarly,
$W_{bb}(0,K_\omega)=\partial_1W_{bb}(0,K_\omega)=\partial_2W_{bb}(0,K_\omega)=0$.
\item
From $W_{ab}(k_0,k)=^b W_{ba}(k_0,-k)=^a W^*_{ba}(-k_0,k)$
we get $W_{ab}(0,K_\omega)=W^*_{ba}(0,K_\omega)$,
$\partial_1W_{ab}(0,K_\omega)=\partial_1W^*_{ba}(0,K_\omega)$,
$\partial_2W_{ab}(0,K_\omega)=\partial_2W^*_{ba}(0,K_\omega)$.
Moreover $W_{ab}(k_0,k)=^d W_{ab}(-k_0,k)$ hence $\partial_0 W_{ab}(0,K_\omega)=0$.

\item
$\partial_0 W_{aa}(k_0,k)=^a-\partial_0 W_{aa}^*(-k_0,-k)=^b-\partial_0 W_{aa}^*(-k_0,k)$
hence is pure imaginary at $k_0=0$; moreover
$\partial_0 W_{aa}(k_0,k)=^c \partial_0 W_{bb}(k_0,-k)=^a-\partial_0 W^*_{bb}(-k_0,k)$
so that $\partial_0 W_{aa}(0,K_\o)=\partial W_{bb}(0,K_\o)=i z$ with $z$ real

\end{enumerate}

\vfill
\eject

\bibliographystyle{amsalpha}

\end{document}